\documentclass[aps,pre,preprint, superscriptaddress]{revtex4-1}
\usepackage{latexsym}
\usepackage{natbib}
\usepackage{amsmath}
\usepackage{amsfonts}
\usepackage{amssymb}
\usepackage[utf8]{inputenc}
\usepackage[T1]{fontenc}
\usepackage[dvips]{color,graphicx}   
\usepackage{verbatim}   
\usepackage{t1enc}
\usepackage{anysize}
\usepackage{tocbibind}
\usepackage{subfig}
\usepackage[percent]{overpic} 
\usepackage{float}  
\usepackage{appendix} 
\usepackage{booktabs} 
\usepackage{natbib}
\usepackage{array}
\usepackage{xcolor}
\usepackage[normalem]{ulem}

\begin{document}
\title{Epidemic control in networks with cliques}
\author{L.  D. Valdez} \affiliation{Instituto de
  Investigaciones F\'isicas de Mar del Plata (IFIMAR)-Departamento de
  F\'isica, FCEyN, Universidad Nacional de Mar del Plata-CONICET, Mar
  del Plata 7600, Argentina.}
\author{L.  Vassallo}\affiliation{Instituto de
  Investigaciones F\'isicas de Mar del Plata (IFIMAR)-Departamento de
  F\'isica, FCEyN, Universidad Nacional de Mar del Plata-CONICET, Mar
  del Plata 7600, Argentina.}
 \author{L.
  A. Braunstein}\affiliation{Instituto de Investigaciones F\'isicas de
  Mar del Plata (IFIMAR)-Departamento de F\'isica, FCEyN, Universidad
  Nacional de Mar del Plata-CONICET, Mar del Plata 7600, Argentina.}
\affiliation{Physics Department, Boston University, Boston, MA 02215,
  United States} \date{\today}

\begin{abstract}
Social units, such as households and schools, can play an important role in controlling epidemic outbreaks. In this work, we study an epidemic model with a prompt quarantine measure on networks with cliques
(a $clique$ is a fully connected subgraph representing a
social unit). According to this strategy, newly infected individuals are detected and quarantined (along with their close contacts) with
probability $f$. Numerical simulations reveal that epidemic outbreaks in networks with cliques
are abruptly suppressed at a transition point $f_c$. However, small
outbreaks show features of a second-order phase transition around
$f_c$. Therefore, our model can exhibit properties of both
discontinuous and continuous phase transitions. Next, we show
analytically that the probability of small outbreaks goes continuously
to 1 at $f_c$ in the thermodynamic limit. Finally, we find that our model exhibits a backward
bifurcation phenomenon.

\end{abstract}

\maketitle

\section{Introduction}

When a novel and dangerous disease unfolds, governments often
implement a wide range of non-pharmaceutical interventions (NPIs) to decrease the burden on health care
services~\cite{flaxman2020estimating,fong2020nonpharmaceutical,markel2007nonpharmaceutical}. These
interventions include, for example, travel bans, quarantine measures,
and school closures. Epidemiological studies have shown that the
spread of contagious diseases depends on multiple factors, including
the network of face-to-face
contacts~\cite{de2004sexual,susswein2020characterizing,luke2007network}. Therefore,
studying the effect of different network structures on the spread of epidemics
becomes essential to develop more effective interventions.

In the last few years, several mathematical models have been proposed
to study NPIs in complex
networks~\cite{wang2015coupled,wang2021literature,kojaku2021effectiveness,rizi2022epidemic}. For
example, St-Onge et al.~\cite{st2021social,st2021master} recently
explored a susceptible-infected-susceptible (SIS) model on networks
with cliques (defined as groups where all members are connected to
each other) and proposed a mitigation strategy that consists of
reducing the maximum clique size. They found that the total fraction of
infected people decreases as the maximum clique size is reduced. Another NPI that
has been extensively studied in the field of complex networks is the
rewiring strategy in which susceptible individuals protect themselves by
breaking their links with infected contacts and creating new ones with
non-infectious people~\cite{gross2006epidemic}. Interestingly, recent
work has shown that this strategy can lead to an explosive epidemic
for a susceptible-infected-recovered (SIR)
model~\cite{durrett2022susceptible,ball2022epidemics}.

Several works have also explored the effect of different quarantine
strategies on the spread of
epidemics~\cite{strona2018rapid,vyasarayani2020complete,hasegawa2017efficiency}. For
example, Hasegawa and Nemoto~\cite{hasegawa2017efficiency}
investigated a susceptible-infected-recovered-quarantined (SIRQ) model
with a ``prompt quarantine strategy'' that works as follows. At each
time step, after individuals become infected, they are immediately
detected with probability $f$, and then the detected ones and their contacts are
placed under quarantine. In that work, they showed (for networks without cliques) that
the probability of an epidemic and the proportion of recovered
people undergo a continuous phase transition. On the other hand,
very recently, B\"orner et al.~\cite{borner2022explosive} studied an
SIRQ model with a quarantine strategy that becomes less effective over
time. More specifically, they considered the case in which the rate at
which individuals are quarantined decreases as the total number of
infected people increases. For a mean-field model (corresponding to a
homogeneously well-mixed population), they showed that the proportion
of recovered people at the final stage could exhibit a discontinuous
transition. However, they also observed that the probability of an
epidemic vanishes continuously around this transition point, so their
model exhibits features of both continuous and discontinuous phase
transitions.

Following the line of research on non-pharmaceutical interventions, in
this paper, we investigate an SIRQ model with a prompt quarantine
strategy on random networks with cliques.  On the one hand, numerical
simulations show that the probability of an epidemic ($\Pi$) vanishes
continuously at a transition point $f=f_c$ (i.e., the probability of a small outbreak, $1-\Pi$, goes to 1 at $f=f_c$). However, numerical simulations also reveal that
the fraction of recovered people ($R$) is abruptly suppressed
around $f=f_c$, so our model displays features of both continuous and
discontinuous phase transitions as in~\cite{borner2022explosive}. Note that this result is markedly
different from the case without cliques, where only a continuous phase
transition was observed~\cite{hasegawa2017efficiency}, as mentioned
above. Finally, we find that our model exhibits the phenomenon of
backward bifurcation. In order to elucidate the origin of these
results, we explore the spread dynamics close to the transition point,
and numerical simulations suggest that the quarantine strategy becomes
less effective over time, which may explain why our model exhibits the
same behavior as in~\cite{borner2022explosive}. 

This paper
is organized as follows. In Sec.~\ref{Sec.Model}, we describe the details of our model. In
Secs.~\ref{Sec.FinS}-\ref{sec.prob}, we investigate the final stage of
an epidemic and the probability of small outbreaks ($1-\Pi$) when only one person is infected at the beginning of the
outbreak. In the
following section, we explore the final stage of an epidemic when a large
proportion of the population is infected at the beginning of the
spreading process. Finally, we present our conclusions.

\section{Model description}\label{Sec.Model}

\subsection{Network with cliques}\label{Sec.Bip}

\begin{figure}[H]
\vspace{0.5cm}
\begin{center}
\begin{overpic}[scale=1.5]{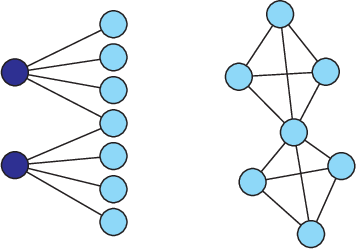}
  \put(0,65){(a)}
  \put(60,65){(b)}
\end{overpic}
\vspace{1cm}
\end{center}
\caption{(Color online) Illustration of a bipartite network (panel a) and its
  projection (panel b). Each blue node represents a clique and each
  light blue node represents an individual.}\label{fig.Bip}
\end{figure}

Networks with cliques can be represented as bipartite networks (as illustrated in Fig.~\ref{fig.Bip}). In this work, we will focus on bipartite networks that are locally tree-like because they have two main advantages. First,
they can be easily generated by using a version of the configuration
model~\cite{molloy1998size,molloy1995critical}, and second, they
simplify the analytical treatment, as explained in~\cite{karrer2010random}.

To generate these networks, we apply the following steps:
\begin{itemize}
\item Step 1) We create two disjoint sets, denoted by $I$ and $C$, where
  $I$ corresponds to the set of individuals and $C$ represents the set of
  cliques. The total numbers of individuals and cliques are denoted by
  $N_I$ and $N_C$, respectively.
\item Step 2) We randomly assign a number $k_I$ of cliques (or
  ``stubs'') to every person according to a probability distribution
  $P(k_I)$. Similarly, we assign a number $k_C$ of individuals (or
  ``stubs'') to every clique according to a probability distribution
  $P(k_C)$. Initially, each stub is unmatched. We denote the total
  number of stubs in sets $I$ and $C$, by $\mathcal{S}_I$ and
  $\mathcal{S}_C$, respectively.  In the limit of large network sizes,
  the relation $\mathcal{S}_C=\mathcal{S}_I$ holds (as explained in~\cite{newman2002spread}). Additionally, in this limit, we have
  that $\mathcal{S}_I=\langle k_I\rangle N_I$ and
  $\mathcal{S}_C=\langle k_C\rangle N_C$, where $\langle k_I\rangle
  =\sum_{k_I} k_IP(k_I)$ and $\langle k_C\rangle =\sum_{k_C} k_C
  P(k_C)$.
\item Step 3) In practice, for finite networks, if $|\mathcal{S}_C-\mathcal{S}_I|<0.01
  \langle k_I\rangle N_I$ then we proceed as follows.  We randomly
  choose one stub from each set and join them together to make a
  complete link (but avoiding multiple connections between individuals
  and cliques). This procedure is repeated until one of these sets is
  empty. On the other hand, if $|\mathcal{S}_C-\mathcal{S}_I|>0.01
  \langle k_I\rangle N_I$, our algorithm returns to Step 1.
\item Step 4) Finally, we eliminate those stubs that remained
  unmatched from the previous step, and project the set of cliques
  onto the set of individuals, as illustrated in Fig.~\ref{fig.Bip}.
\end{itemize}

\subsection{Susceptible-infected-recovered-quarantined model}\label{Sec.SIRQ}

Let us first introduce the susceptible-infected-recovered model (SIR),
and some definitions.

The SIR model splits the population into three compartments called
susceptible ($S$), infected ($I$), and recovered ($R$). Here, the symbols $S$,
$I$, and $R$ refer to both the state of an individual and the proportion of
the population in each compartment, where $S+I+R=1$. For a
discrete-time SIR model, all individuals synchronously update their
states according to the following rules. At each time step, $t\to t+1$,
every infected individual
\begin{enumerate}
\item transmits the disease to each susceptible
  neighbor with probability $\beta$,
\item recovers from the disease after
  being infected for $t_r$ time steps (which is called the recovery
  time) and becomes permanently immune. In this paper, we will use $t_r=1$.
\end{enumerate}
Typically, the spreading process starts with a single infected individual,
called the ``index-case'', and the rest of the population is
susceptible. The disease then spreads through the population until the
system reaches a final stage with only susceptible and recovered
individuals. If the disease dies out after a few time steps and only
an insignificant fraction of the population has become infected, then
such an event is defined as a small outbreak. Conversely, the outbreak
turns into an epidemic if the fraction of recovered people is
macroscopic at the final stage. In the last few years, several works
have also studied the case in which a macroscopic fraction $I_0$ of
the population is infected at the beginning of the spreading
process~\cite{radicchi2020epidemic,miller2014epidemics,krapivsky2021infection,machado2022effect,hasegawa2016outbreaks}. This
case is usually referred to as a non-trivial or large initial
condition.

A widely used measure to predict,
whether a disease will develop into a small outbreak or an epidemic is
the basic reproduction number $R_0$, defined as the average number of
secondary cases infected by the
index-case~\cite{brauer2008mathematical}. For a value of $R_0$ less
than 1, the probability of a disease becoming an epidemic is known
to be zero ($\Pi=0$), while for $R_0$ greater than 1, this
probability is positive ($\Pi>0$). Finally, around $R_0=1$, there is a
second-order phase transition where many quantities behave as
power-laws~\cite{stauffer2014introduction,grassberger1983critical}. For
example, at $R_0=1$ the probability distribution of the number of
recovered individuals for small outbreaks, denoted by $P(s)$, decays
algebraically as $P(s)\sim s^{-(\tau-1)}$, where $\tau$ is called the
Fisher exponent~\cite{stauffer2014introduction}.

As explained in the Introduction, an extension of the SIR model that was proposed in~\cite{hasegawa2017efficiency}, introduces
a $Q$ compartment in order to study the effect of a prompt quarantine
strategy on the epidemic spread. In this model, the states of the
nodes were updated asynchronously. However, in our work, we will
consider a synchronous version of that model in order to simplify the
analytical study. More precisely, our model works as follows: at time
$t$,
\begin{enumerate}
\item All infected individuals are detected and isolated with probability $f$,
  i.e., they move to the $Q$ compartment.
\item Next, all the neighbors of the individuals who were
  isolated in the previous step, also move to the $Q$ compartment.
\item After that, those infected individuals who have not been isolated, will transmit the disease to each
  susceptible neighbor with probability $\beta$.
\item Finally, individuals who are still in the $I$ compartment and
  have been infected for $t_r$ time steps, will recover~\cite{Exxon01}. Likewise,
  people who have been infected and then quarantined will move to the
  $R$ compartment after $t_r$ time steps, so $R$ represents the
  proportion of the population ever infected. In this
paper, we present results only for $t_r=1$; however, we have
verified that our findings remain qualitatively unchanged for $t_r>1$
(not shown here).
\end{enumerate}
Similarly to the standard SIR model, at the final stage of the SIRQ
model, the population consists solely of susceptible, recovered and
quarantined individuals. 

Note that, according to the rules of our model, it is sufficient to
detect a single infected person in a clique to quarantine the entire
clique.  Therefore, larger (smaller) cliques have a higher (lower)
probability of being quarantined. On the other hand, from one perspective, our model could be seen as a spreading process in higher-order networks~\cite{battiston2020networks,battiston2021physics}
because the transition from a susceptible to a quarantined state is
not caused by pairwise interactions but rather by group
interactions. Typically, in models with higher-order structures, nodes become "infected" through group interactions, and after that, they transmit the "infection" to other nodes. However, it should be noted that in our model, quarantined individuals are removed from the system, so they cannot transmit their state to the rest of the population, unlike other contagion models with higher-order structures.

In the following sections, we will study our SIRQ model on networks
with cliques.

\section{Results}\label{Sec.Res}
\subsection{Final stage}\label{Sec.FinS}
In this section, we investigate the final stage of the SIRQ model for
random regular (RR) networks with cliques, defined as networks in
which every clique has $k_C$ members and every individual belongs to
$k_I$ cliques. We will show numerical results for RR with
$k_I=3$, and $k_C=7$ and focus only on the case
where a single individual is infected at the beginning of the dynamic
process. In Appendix~\ref{Sec.AppAddit}, we present additional results
for networks in which $k_C$ and $k_I$ follow other
probability distributions.

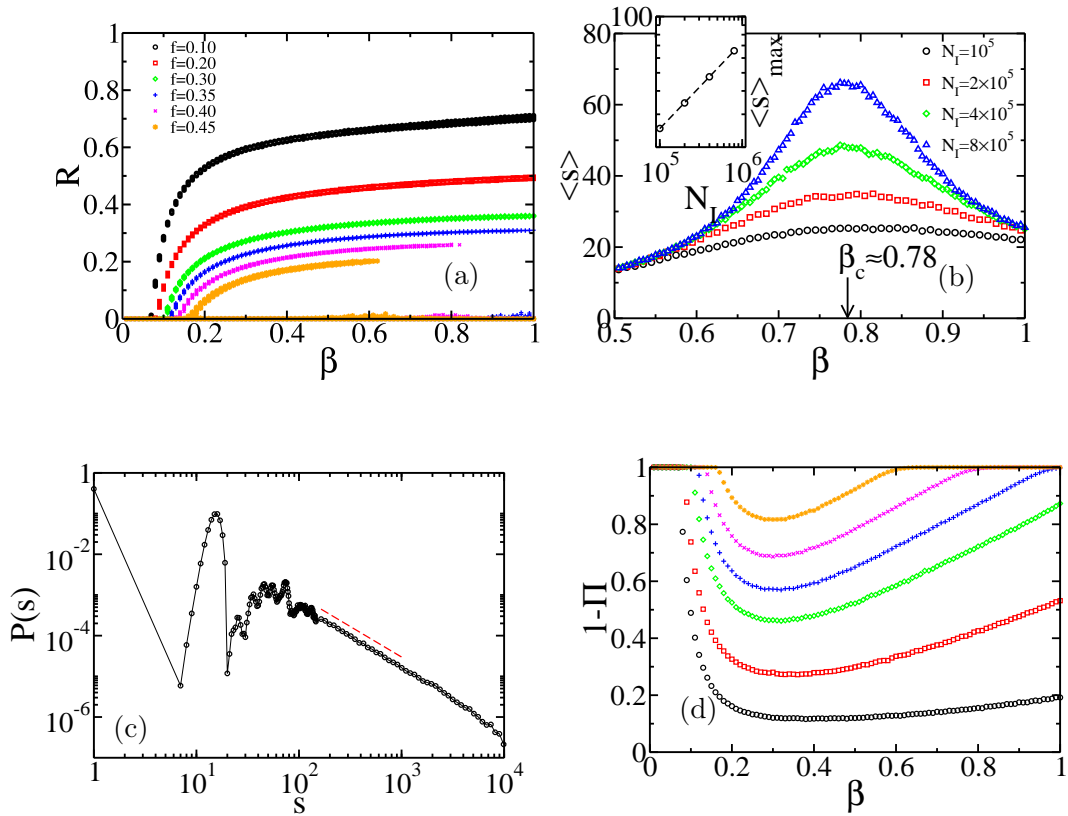
\begin{figure}[H]
\vspace{0.5cm}
\begin{center}
\begin{overpic}[scale=0.25]{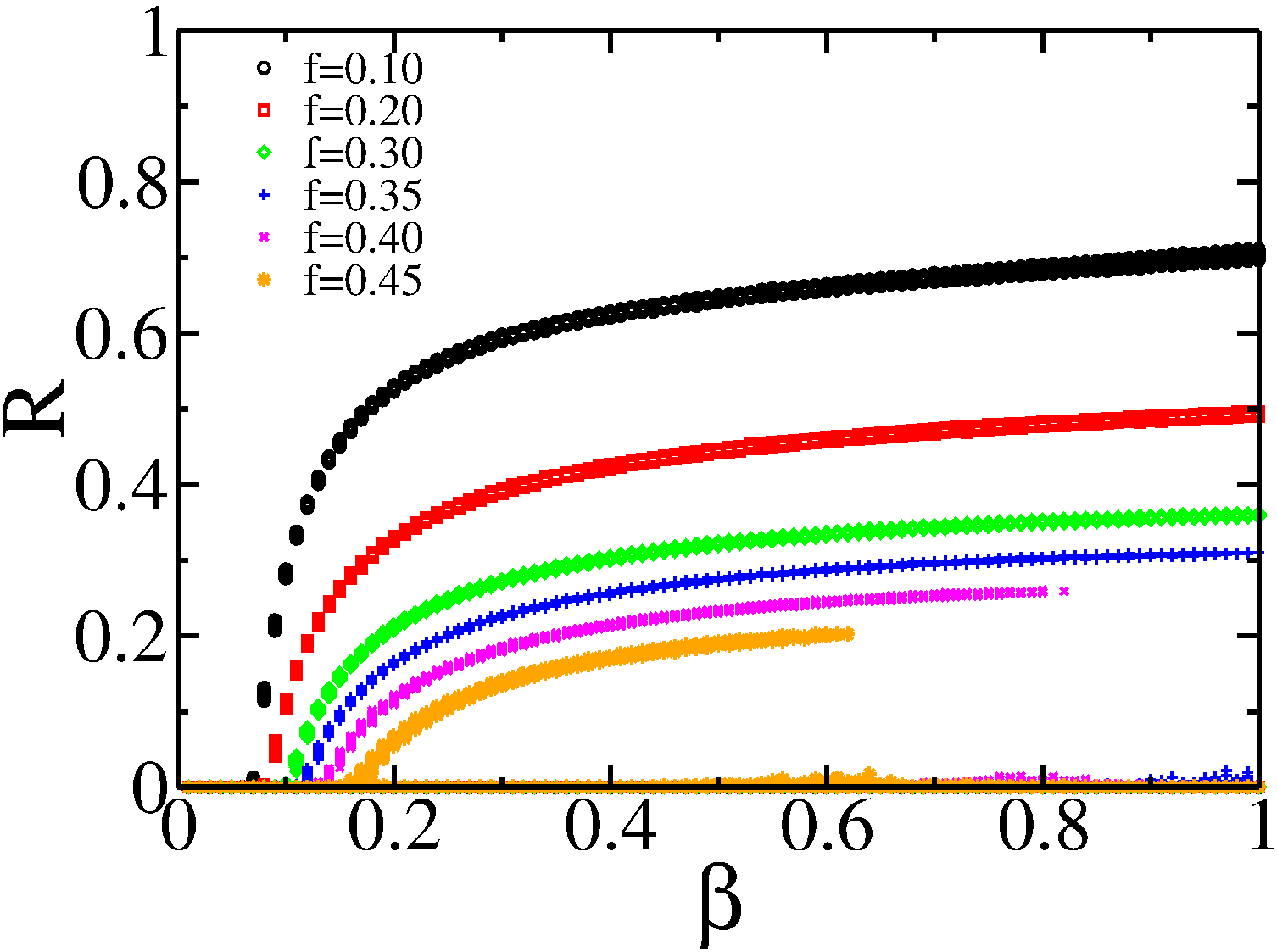}
  \put(80,20){(a)}
\end{overpic}
\vspace{0.5cm}
\vspace{0.5cm}
\begin{overpic}[scale=0.25]{Fig02b.eps}
  \put(80,20){(b)}
\end{overpic}
\vspace{0.0cm}
\begin{overpic}[scale=0.25]{Fig02c.eps}
  \put(20,15){(c)}
\end{overpic}
\hspace{0.5cm}
\begin{overpic}[scale=0.25]{Fig02d.eps}
  \put(20,20){(d)}
\end{overpic}
\vspace{1cm}

\vspace{0.0cm}

\end{center}
\caption{(Color online) Panel a: Scatter-plot of the fraction of
  recovered people at the final stage, $R$, as a function of
  $\beta$ for a RR network with $k_C=7$, $k_I=3$, $N_I=10^6$, and
  different values of the probability of detection $f$. Results were
  obtained from $10^3$ stochastic realizations. Panel b: $\langle s\rangle$ against $\beta$ for
  $f=0.4$ and several values of $N_I$. Results were averaged over
  $10^5$ realizations. The vertical arrow indicates the peak position $\beta_c$ of $\langle s\rangle$ for $N_I=8\times 10^5$. In the inset, we show the height of the peak of
  $\langle s \rangle$, which we call $\langle s \rangle_{max}$ (in
  log-log scale) for the same values of $N_I$ as in the main plot. The
  dashed line corresponds to a power-law fit with an exponent of 0.46. Panel
  c: distribution $P(s)$ for $\beta=0.78$, $f=0.4$, and $N_I=10^6$,
  obtained from $3\times 10^5$ stochastic realizations (symbols). The solid
  black line is a guide to the eye, and the dashed red line is a
  power-law function with an exponent equal to $\tau-1=1.5$. Panel d: Probability
  of a small outbreak, $1-\Pi$, against $\beta$ for the same parameter values
  as in panel a.  Results were averaged over $10^5$ stochastic
  realizations.}\label{fig.Pinf}
\end{figure}

In Fig.~\ref{fig.Pinf}a, we show a scatter-plot of $R$ vs. $\beta$ for
several values of the probability of detection $f$. For low values of
$f$, we observe that the transition from an epidemic-free phase to an
epidemic phase is continuous. However, for $f\gtrsim 0.35$, we see
that as $\beta$ increases, another phase transition exists above which
the fraction of recovered individuals is abruptly suppressed. This
transition is also observed in other network topologies (see
Appendix~\ref{Sec.AppAddit}), especially in networks containing larger
cliques. In Sec.~\ref{Sec.back}, we will show that
around this transition point, a backward bifurcation occurs.

To delve deeper into the nature of the transition point at which $R$
is abruptly suppressed, we will study how small outbreaks behave
around this point. Here, we consider that a small outbreak occurs when the fraction of the recovered people is below 1\% at the final stage. Fig.~\ref{fig.Pinf}b shows the average number of
recovered individuals for small outbreaks $\langle s\rangle$ vs. $\beta$ for
$f=0.4$. Interestingly, we note that $\langle s\rangle$ exhibits a
peak around a value of $\beta$ that we call $\beta_c$, which roughly
corresponds to the point at which $R$ is abruptly suppressed (see
Fig.~\ref{fig.Pinf}a). Furthermore, the height of this peak increases
with $N_I$ as a power-law (see the inset of Fig.~\ref{fig.Pinf}b ),
which is a typical finite-size effect of a second-order phase
transition~\cite{stauffer2014introduction}. On the other hand,
Fig.~\ref{fig.Pinf}c shows the probability distribution of the number
of recovered individuals for outbreaks at $\beta=\beta_c$. It can be
seen that $P(s)$ decays as a power-law. Finally, in
Fig.~\ref{fig.Pinf}d, we display the probability of a small outbreak,
$1-\Pi$, as a function of $\beta$ (note that $\Pi$ is the
probability that an epidemic occurs), and we get that $1-\Pi$ goes
continuously to 1 around $\beta=\beta_c$, which again is a feature of other epidemic and percolation models in random networks
with a continuous phase
transition~\cite{kenah2007second,meyers2006predicting}. Therefore, if
we take together the results of Figs.\ref{fig.Pinf}b-d, they all
suggest that quantities associated with small outbreaks will exhibit properties of a continuous phase transition.

\begin{figure}[H]
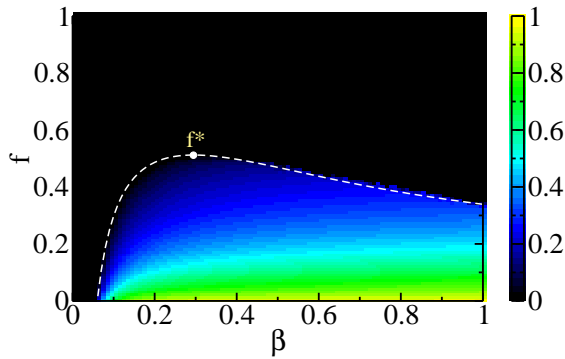

\vspace{0.5cm}
\begin{center}
\begin{overpic}[scale=0.25]{Fig03.eps}
  \put(70,20){}
\end{overpic}
\vspace{1cm}
\vspace{0.0cm}
\end{center}
\caption{(Color online) Heat-map of $R$ in the plane $\beta-f$ for RR networks with cliques (with $k_C=7$ and $k_I=3$), obtained from numerical
  simulations. To compute $R$, we took into account only
  those realizations in which an epidemic occurs ($R>1$\%).  Darker
  colors represent a low value of $R$ (black corresponds to $R=0$) and
  brighter colors a higher value of $R$ (yellow corresponds to
  $R=1$). Simulation results were averaged over $10^3$ stochastic
  realizations with $N_I=10^5$. The dashed white line was obtained from
  Eq.~(\ref{eq.R0}) for $R_0=1$, and the point $f^*=0.51$ corresponds to
  the value of $f$ above which the system is in an epidemic-free phase
  for any value of $\beta$.}\label{fig.Phase}
\end{figure}

To provide a broader picture of the effect of our strategy on networks
with cliques, in Fig.~\ref{fig.Phase}, we show the heat-map of $R$ when an epidemic occurs in the plane
$\beta-f$. From this figure, we observe that there is a minimum detection
probability $f^*$, above which the system is always in an
epidemic-free phase. On the other hand, we also find that in the
region $\beta\lesssim 1$, an abrupt color change occurs around
$f\approx 0.4$, which indicates that the system undergoes a
discontinuous transition in that region of the parameter space.

Next, we will compute the
basic reproduction number, $R_0$.  As mentioned in
Sec.~\ref{Sec.SIRQ}, $R_0$ is a widely used quantity to predict
whether a disease outbreak will become an epidemic or die out
quickly, and typically, around $R_0=1$, a second-order phase transition occurs. In order to estimate $R_0$, we adapt the approach
proposed in~\cite{miller2009spread}, leading us to the
following expression for RR networks with cliques:
\begin{eqnarray}\label{eq.R0}
R_0&=&\frac{\epsilon_1+\epsilon_2}{\beta(k_C-1)},
\end{eqnarray}
with
\begin{eqnarray}
  \epsilon_1&=&(k_C-1)(1-\beta)\left[(\beta(1-f)+(1-\beta))^{k_C-2}-(\beta(1-\beta)(1-f)+(1-\beta))^{k_C-2}\right],\\
  \epsilon_2&=&(1-f)(1-\beta f)^{k_C-2}(k_I-1)(k_C-1)^2\beta^2.
\end{eqnarray}
In Eq.~(\ref{eq.R0}):
\begin{enumerate}
\item the denominator is the average number of individuals (within a clique) who
  are infected by the index-case. We refer to these individuals as the ``first generation.''
\item the numerator corresponds to the average number of people who
  are infected by the first generation. In Appendix~\ref{Sec.AppR0},
  we explain how to derive the expressions of $\epsilon_1$ and
  $\epsilon_2$.
\end{enumerate}
In Fig.~\ref{fig.Phase}, we plot the set of points ($\beta_c,f_c$) that satisfy the constraint $R_0=1$. In particular, for $\beta_c=1$, it
can be easily obtained from Eq.~(\ref{eq.R0}) that $f_c$ is given by
\begin{eqnarray}\label{eq.fcR0}
f_c=1-\left(\frac{1}{(k_I-1)(k_C-1)}\right)^{\frac{1}{k_C-1}}.
\end{eqnarray}
Remarkably, from Fig.~\ref{fig.Phase}, we can see that the predicted
curve agrees well with the entire boundary between the
epidemic and epidemic-free phases, including in the region where a
discontinuous transition occurs. In Sec.~\ref{Sec.back}, we will
see that this result is consistent with a backward bifurcation phenomenon around
$R_0=1$.

In summary, in this section we found that the
SIRQ model on networks with cliques has a discontinuous phase
transition, but at the same time, several quantities (specifically, $\langle s \rangle$,
$1-\Pi$, and $P(s)$) display the same features of a
continuous phase transition. We note, however, that the results shown in this section were
obtained from simulations in finite networks and from approximate
formulas. In the following section, we will demonstrate that in the
thermodynamic limit ($N_I\to \infty$), the probability of a small outbreak
($1-\Pi$) goes continuously to 1 at the transition point for $\beta=1$.

\subsection{Probability of a small outbreak}\label{sec.prob}

In this section, we will describe the SIRQ model as a forward
branching process~\cite{valdez2020epidemic,kenah2007second} 
to calculate the probability of a small outbreak $1-\Pi$ and the transition
point $f=f_c$ for $\beta=1$. Here, we focus only on RR networks with cliques,
but in Appendix~\ref{Sec.AppPi}, we compute these quantities for
other network structures.

Branching theory has been extensively applied to the study of many
processes on random networks, including cascading
failures~\cite{buldyrev2010catastrophic,valdez2020cascading}, disease
transmission~\cite{newman2002spread,pastor2015epidemic,
  wang2018critical}, random
percolation~\cite{dong2021optimal,cohen2002structural}, k-core
percolation~\cite{baxter2011heterogeneous,di2019insights} and fractional
percolation~\cite{shang2014vulnerability,valdez2022emergent}. For
an SIR model, this theory was first applied to networks without
cliques to calculate the behavior of various quantities as a function
of $\beta$~\cite{newman2002spread,pastor2015epidemic}. Later on,
multiple works used branching theory to study the SIR model on
networks with
cliques~\cite{mann2021random,mann2021exact,allard2012bond,gleeson2009bond}. However,
their calculations were usually more complex because they required an
exhaustive enumeration of transmission events occurring within a clique with at
least one infected person.  But, for $\beta=1$, these calculations can be
substantially simplified. This is because when
individuals become infected (in a clique composed of susceptible
members), at the next time step, they will transmit the disease to
the rest of the clique members with probability 1, unless an
intervention strategy is applied. Therefore, in what follows, we will
focus only on the case $\beta=1$.

To compute the probability of a small outbreak
$1-\Pi$, we first need to calculate the probability
$\phi$ that an infected individual (reached through a link) will not
generate an epidemic~\cite{kenah2007second,meyers2006predicting}.  By using the branching process approach, it can
be found that $\phi$ is the solution of the following self-consistent
equation:
\begin{eqnarray}\label{eq.phi}
  \phi&=&\left[ ((1-f)\phi)^{k_C-1}+1-(1-f)^{k_C-1}\right]^{k_I-1}.
\end{eqnarray}

The l.h.s of this equation is the probability that an infected
individual ``$j$'' reached through a link, does not cause an
epidemic. On the other hand, the r.h.s. is the probability
that an infected individual ``$j$'' transmits the disease, but none of
the $k_I-1$ outgoing cliques will be able to cause an epidemic. This is
because one of the following two events happens to every clique:
\begin{enumerate}
\item with probability $1-(1-f)^{k_C-1}$, at least one member (other than ``$j$'') is detected, so the whole clique is placed under quarantine,
\item with probability $((1-f)\phi)^{k_C-1}$, none of its members are detected but also they will not be able to generate epidemics.
\end{enumerate}

After solving Eq.~(\ref{eq.phi}), the probability $1-\Pi$ that an
index-case does not cause an epidemic can be obtained from the equation
\begin{eqnarray}\label{eq.noPi}
1-\Pi&=&f+(1-f)\left[((1-f)\phi)^{k_C-1}+1-(1-f)^{k_C-1}\right]^{k_I},
\end{eqnarray}
where, in the r.h.s. :
\begin{enumerate}
\item the first term corresponds to the probability that the index-case is detected
\item the second term corresponds to the scenario where the index-case is
  not detected and transmits the disease, but none of the $k_I$
  outgoing cliques will be able to generate an epidemic, similarly to Eq.~(\ref{eq.phi}).
\end{enumerate}

It is worth noting that Eqs.~(\ref{eq.phi})-(\ref{eq.noPi}) are valid only if the initial fraction of index-cases is
infinitesimal.

Another quantity of interest that can be calculated in the limit of large network
sizes is the critical threshold $f_c$ at which a phase transition
occurs. To derive $f_c$, we take derivatives of both sides of Eq.~(\ref{eq.phi}) at
$\phi=1$ and obtain:
\begin{eqnarray}\label{eq.fcBranch}
f_c=1-\left(\frac{1}{(k_I-1)(k_C-1)}\right)^{\frac{1}{k_C-1}},
\end{eqnarray}
which has the same expression as in Eq.~(\ref{eq.fcR0}).

\begin{figure}[H]
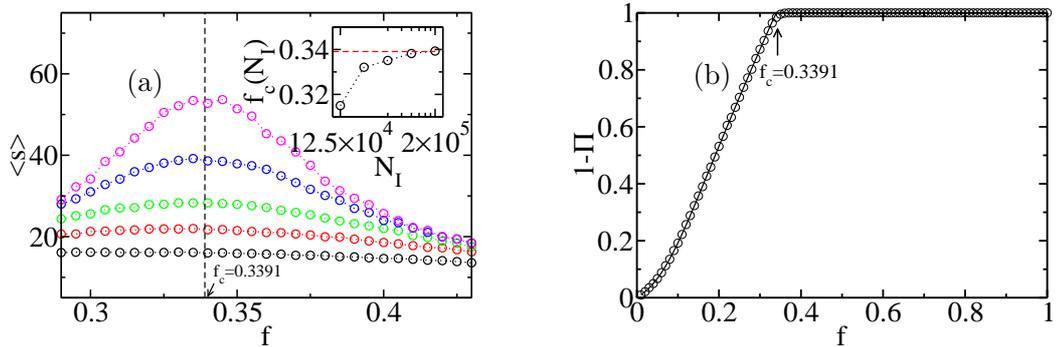

\vspace{0.5cm}
\begin{center}
\begin{overpic}[scale=0.25]{Fig04a.eps}
  \put(25,55){(a)}
\end{overpic}
\hspace{1cm}
\begin{overpic}[scale=0.25]{Fig04b.eps}
  \put(25,55){(b)}
\end{overpic}
\vspace{1cm}
\end{center}
\caption{(Color online) Panel a: $\langle s\rangle$ vs.
  $f$ for $\beta=1$ and RR networks with cliques ($k_C=7$ and $k_I=3$)
  for different network sizes (from bottom to top:
  $N_I=1.25\times 10^4,\;2.5\times 10^4,\;5\times 10^4,\;10^5$, and $2\times 10^5$). Symbols
  correspond to simulation results averaged over $10^5$ stochastic realizations. The
  vertical dashed line indicates the predicted value of $f_c$ obtained
  from Eq.~(\ref{eq.fcBranch}). In the inset, we show the peak position of $\langle s \rangle$ (estimated from the main
  plot), called $f(N_I)$, as a function of $N_I$ in a linear-log scale. The dashed line
  corresponds to our theoretical prediction of $f_c$. Dotted lines are
  a guide to the eye. Panel b: Probability of a small outbreak
  ($1-\Pi$) vs. $f$  for $\beta=1$
  and RR networks with cliques ($k_C=7$ and $k_I=3$). The line
  corresponds to the theory given by
  Eqs.~(\ref{eq.phi})-(\ref{eq.noPi}), and symbols are
   simulation results averaged over $10^5$ realizations with $N_I=10^6$.}\label{fig.TheorT1}
\end{figure}

To verify the validity of our theoretical analysis, we performed
numerical simulations of the SIRQ model on RR networks with
cliques. In Fig.~\ref{fig.TheorT1}a, we show the mean size of small
outbreaks $\langle s\rangle$ vs. $f$ for 
different network sizes ($N_I$). It can be seen that as $N_I$
increases, the peak position of $\langle s\rangle$ (that we
call $f_c(N_I)$) converges to the critical threshold $f_c$ predicted
by Eq.~(\ref{eq.fcBranch}). On the other hand, in
Fig.~\ref{fig.TheorT1}b, we display the probability of a small outbreak,
$1-\Pi$, obtained from our simulations and theoretical predictions [see
  Eqs.~(\ref{eq.phi})-(\ref{eq.noPi})]. As seen in this figure, the agreement
between theory and simulation is excellent. In addition, we observe
that $1-\Pi$ goes continuously to 1 (i.e., $\Pi\to 0$) at the critical
threshold $f=f_c$ predicted by Eq.~(\ref{eq.fcBranch}). Thus, our findings in this section provide further evidence that small outbreaks display features of a continuous phase
transition around $f=f_c$, as noted in the previous
section. 

In the next section, we will investigate the effect of non-trivial
initial conditions on the final stage of the propagation process and
discuss the mechanism leading to the discontinuous transition observed
in Sec.~\ref{Sec.FinS}.

\subsection{Backward bifurcation}\label{Sec.back}
In previous sections, we focused our attention only on the case where a
single index-case was infected at the beginning of the outbreak. Here, we will study the effect of a non-trivial initial condition
on the final stage of the propagation process. To this end, we conduct
numerical simulations in which the fraction of infected individuals
at $t=0$ (denoted by $I_0$) is macroscopic. In particular, we are
interested only in the case where $f>f_c$ (i.e., $R_0<1$) because for $f<f_c$
(i.e., $R_0>1$), we have already found that an epidemic can take off
even from a single index-case (see Sec.~\ref{sec.prob}).

\begin{figure}[H]
\vspace{0.5cm}
\begin{center}
\begin{overpic}[scale=0.35]{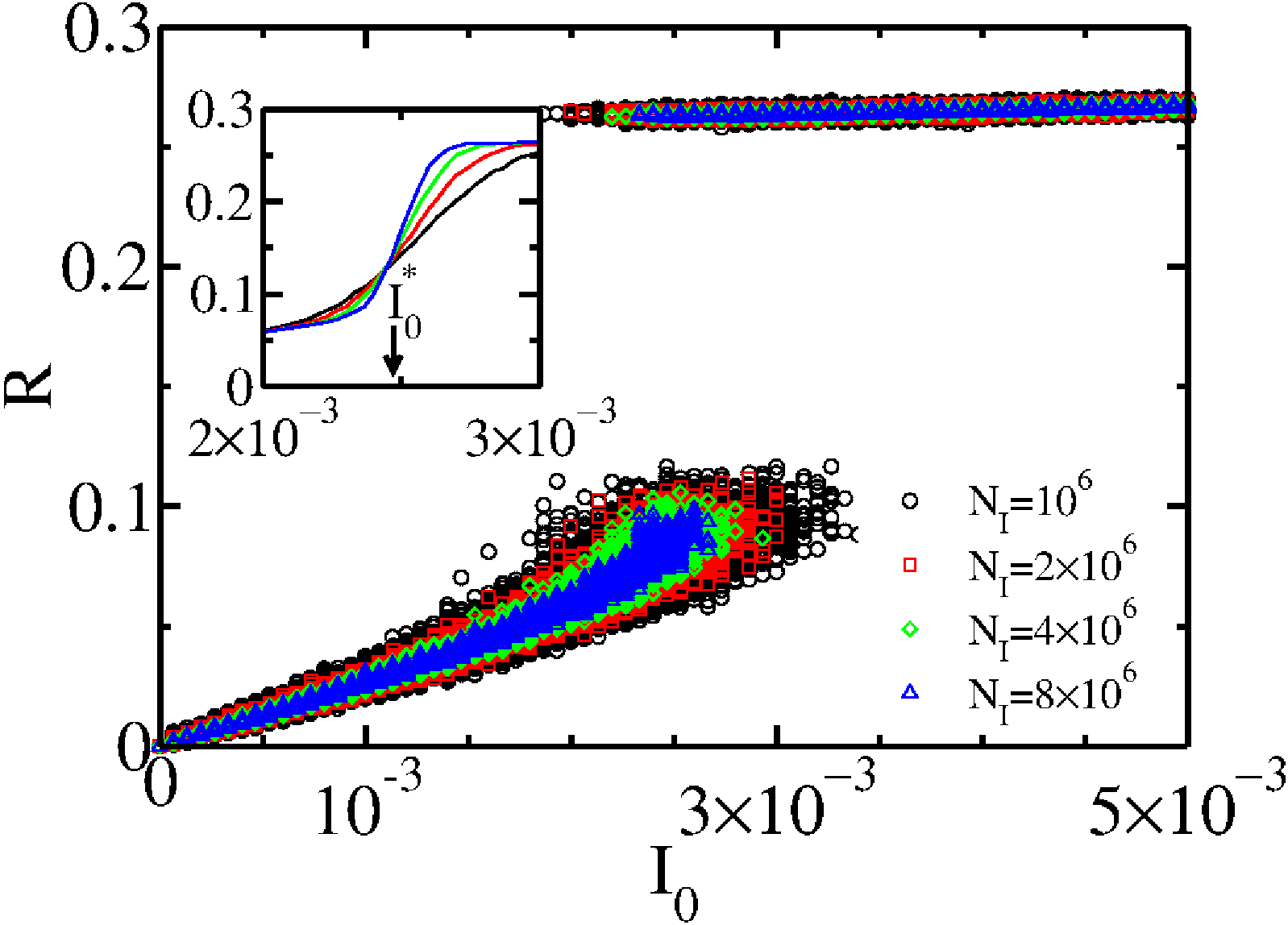}
  \put(80,55){}
\end{overpic}
\vspace{0.5cm}
\vspace{0.0cm}
\end{center}
\caption{(Color online) Scatter-plot of $R$ vs. $I_0$
  obtained from numerical simulations for $\beta=1$ and $f=0.40$ in a
  RR network with $k_C=7$, $k_I=3$, and different network sizes
  $N_I$. Inset: Average value of $R$ as a function of $I_0$ for the
  same parameter values used in the main plot. Numerical results were averaged
  over $10^4$ stochastic realizations. The vertical arrow indicates
  the value of $I_0^*$ around which $R$ undergoes a phase
  transition.}\label{fig.I0}
\end{figure}

Fig.~\ref{fig.I0} shows a scatter-plot of the proportion of recovered
people $R$ at the final stage as a function of $I_0$ for
$\beta=1$ and $f=0.40$ (which is greater than $f_c=0.3391$; see
Sec.~\ref{sec.prob}), and for several network sizes
$N_I$. Additionally, in the inset, we plot the average value of $R$
vs. $f$ for the same parameter values used in the main
plot. Interestingly, we obtain that $R$ has an abrupt jump around
$I_0\approx 2.5\times 10^{-3}\equiv I_0^*$. Therefore, our numerical
simulations reveal that the final fraction of recovered people strongly
depends on the initial fraction of infected individuals for
$R_0<1$. In the language of bifurcation theory, these findings imply
that our model undergoes a backward
bifurcation~\cite{gumel2012causes}, i.e., the final fraction of
recovered people is bistable for $R_0<1$ ($f>f_c$). In
Appendix~\ref{sec.appBif}, we present additional results showing that
the system is also bistable for other values of $f$ and network topologies.

Previous studies have shown that this type of bifurcation can be
caused by multiple mechanisms, such as exogenous re-infection and
the use of an imperfect vaccine against
infection~\cite{gumel2012causes}. On the other hand, very recently,
B{\"o}rner et al.~\cite{borner2022explosive} proposed a mean-field
SIRQ model to explore different quarantine measures whose
effectiveness decreases over time. Although not explicitly mentioned
in that work, it can be seen that their model is sensitive to initial
conditions for $R_0<1$. Thus, a backward bifurcation phenomenon can
also be caused by a quarantine measure that becomes less effective
over time.  Additionally, in~\cite{borner2022explosive}, it was
shown that a discontinuous epidemic phase transition occurs, and the
probability of an epidemic vanishes around the transition point.

\begin{figure}[H]
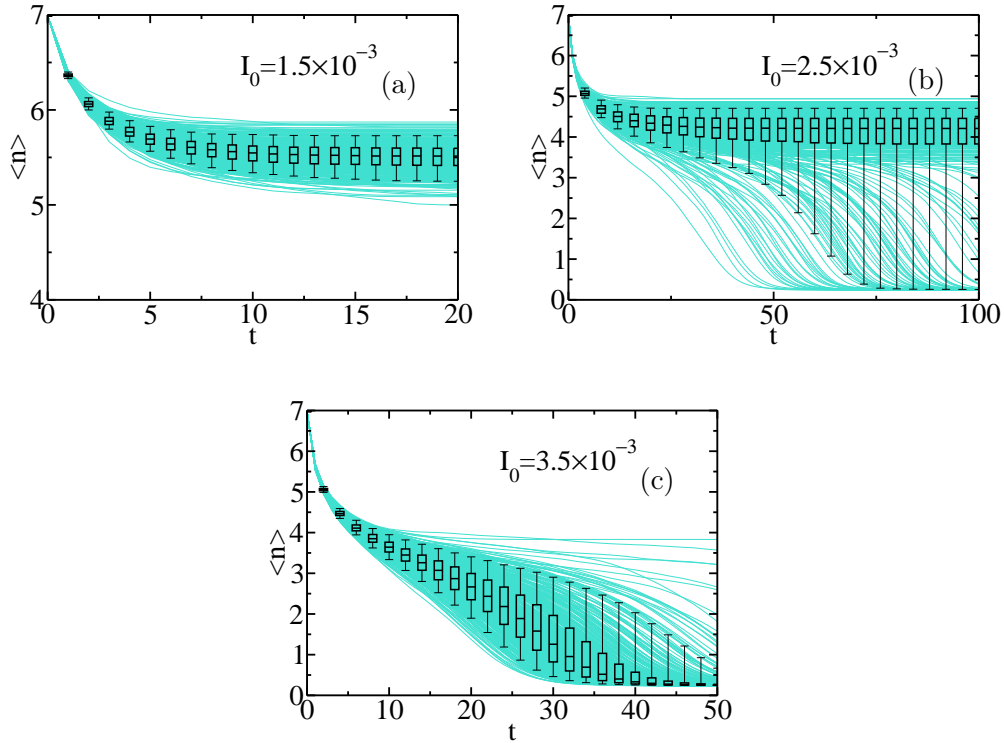

\vspace{0.5cm}
\begin{center}
\begin{overpic}[scale=0.25]{Fig06a.eps}
  \put(80,55){(a)}
\end{overpic}
\vspace{0.7cm}
\hspace{0.5cm}
\begin{overpic}[scale=0.25]{Fig06b.eps}
  \put(80,55){(b)}
\end{overpic}
\vspace{0.5cm}
\begin{overpic}[scale=0.25]{Fig06c.eps}
  \put(80,55){(c)}
\end{overpic}
\vspace{0.0cm}
\end{center}
\caption{(Color online) Time evolution of the average number of
  individuals (either in a susceptible or infected state) in a clique,
  denoted by $\langle n \rangle$, for $f=0.40$, $\beta=1$, and
  several initial conditions: $I_0=1.5\times 10^{-3}$ (panel a),
  $I_0=2.5\times 10^{-3}$ (panel b), and $I_0=3.5\times 10^{-3}$ (panel
  c). We generated 500 simulation trajectories (light blue lines) on
  RR networks with $k_C=7$, $k_I=3$, and $N_I=10^6$. Box plots show the
  5th, 25th, 50th, 75th and 95th percentile values of $\langle
  n\rangle$.}\label{fig.evoln}
\end{figure}

To explain why our model is sensitive to initial conditions
for $R_0<1$, we will next measure the time evolution of $\langle
n\rangle$ for several values of $I_0$, where $\langle n \rangle$ is
defined as the average number of members (either in a susceptible or
infected state) in a clique. In particular, for RR networks with
cliques, the inequality $\langle n\rangle \leq k_C$ holds. From
Fig.~\ref{fig.evoln}, we can clearly see that $\langle n\rangle$ is a
decreasing function with time, or in other words, cliques become
smaller as the population moves into the $Q$ and $R$
compartments. This leads us to the conclusion that the effectiveness
of our strategy diminishes over time (as in~\cite{borner2022explosive}) because, as indicated in
Sec.~\ref{Sec.SIRQ}, smaller cliques are less likely to be placed
under quarantine. Therefore, based on what was observed in~\cite{borner2022explosive}, we conjecture that  a decrease in $\langle n \rangle$ over
time could explain why our model displays an abrupt transition and a
backward bifurcation diagram, as seen in Secs.~\ref{Sec.FinS}
and~\ref{Sec.back}, respectively.

\section{Conclusions}

In summary, in this paper, we have investigated an SIRQ model
with a prompt quarantine measure on networks with cliques. Numerical
simulations revealed that epidemics could be abruptly suppressed at
a critical threshold $f_c$, especially on networks with larger cliques
(as shown in Appendix~\ref{Sec.AppAddit}). In contrast, we observed that  small outbreaks exhibit properties of a continuous
phase transition around $f_c$. Furthermore, using branching theory, we demonstrated
that the probability of a small outbreak goes continuously to 1 at
$f=f_c$ for $\beta=1$. Therefore, these results indicate that our model can exhibit features of both
continuous and discontinuous transitions. Next, we explored the impact
of a macroscopic fraction of infected population at the beginning of
the epidemic outbreak, and found that for $R_0<1$, a backward
bifurcation phenomenon emerges. Finally, numerical simulations showed
that the quarantine measure becomes less effective over time, which
could explain why our model exhibits an abrupt transition and a
backward bifurcation phenomenon.

Several lines of research can be derived from this work. For example,
one question that remains open is whether the fraction of
recovered people (in the event of an epidemic) can be predicted by
branching theory since in this paper we have only used this theory to study small outbreaks. On the other hand, our model could be extended to
include a time-lag between infection and detection. Another relevant modification would be to allow quarantined individuals to return to
the network after a certain period of time (especially those who were
susceptible) because it is unrealistic to assume that they will remain
isolated until the end of an epidemic outbreak. Additionally, a natural extension of our work would be to study the phenomenon of mesoscopic localization~\cite{st2021social,st2021master}. Lastly, our model could
be studied in higher-order networks with simplicial complexes. It is known that simplicial contagion
models can lead to explosive epidemic transitions~\cite{battiston2021physics,battiston2020networks}, so it would be interesting to
investigate how they compete with a prompt quarantine measure.  We
will explore some of these extensions in a forthcoming work.

\section*{Acknowledgement}
We thank UNMdP (EXA 956/20), FONCyT (PICT 1422/2019) and CONICET, Argentina, for financial support. We also thank Dr. C. E. La Rocca and Lic. I. A. Perez for valuable discussions.

\appendix

\section{Basic reproduction number for RR networks with cliques}\label{Sec.AppR0}

In~\cite{miller2009spread}, Miller estimated the basic reproduction
number $R_0$ for random networks with cliques using the concept of
rank proposed by Ludwing~\cite{ludwig1975final}. He found that $R_0$ can be well estimated by the
following expression
\begin{eqnarray}
R_0=\frac{\langle N_2 \rangle}{\langle N_1 \rangle},
\end{eqnarray}
where $\langle N_1 \rangle$ and $\langle N_2 \rangle$ are the average number of infected people of
rank 1 and 2, respectively.

In our work, we use a similar approach to the one proposed in~\cite{miller2009spread}, but estimate $R_0$ as the following ratio:
\begin{eqnarray}
R_0=\frac{\langle \mathcal{N}_2 \rangle}{\langle \mathcal{N}_1 \rangle},
\end{eqnarray}
where:
\begin{itemize}
\item $\langle \mathcal{N}_1 \rangle$ is the average number of people (within a clique)  infected by the index-case, that we call the ``first generation'',
\item $\langle \mathcal{N}_2\rangle$ is
the average total number of individuals infected by the people from the first
generation.
\end{itemize}

In what follows, we derive the expressions of $\langle \mathcal{N}_1\rangle$ and $\langle \mathcal{N}_2\rangle$ for the case of
random RR networks in which every clique has $k_C$ members, and each
person belongs to $k_I$ cliques.

\subsection{Derivation of $\langle \mathcal{N}_1\rangle$}

Let us consider that, at time $t=0$, there is a single index-case and the rest
of the population is susceptible. If we assume that the index-case is not
detected, it follows that the probability that the index-case will transmit the disease to $\mathcal{N}_1$
individuals (in a clique with $k_C$ members) is given by,
\begin{eqnarray}\label{Eq.N1}
P(\mathcal{N}_1)=\binom{k_C-1}{\mathcal{N}_1}\beta^{\mathcal{N}_1}(1-\beta)^{k_C-1-\mathcal{N}_1}.
\end{eqnarray}
Then, the average number of people infected by the
index-case at $t=1$ is
\begin{eqnarray}
  \langle \mathcal{N}_1 \rangle&=&\sum_{\mathcal{N}_1=0}^{k_C-1}\mathcal{N}_1P(\mathcal{N}_1),\nonumber\\
  &=&(k_C-1)\beta.\label{Eq.N1A}
\end{eqnarray}
Note that the people who get infected at this time step are at a chemical distance of $\ell=1$ from
the index-case.

\subsection{Derivation of $\langle \mathcal{N}_2\rangle$}

After the index-case has infected $\mathcal{N}_1$ people, one of the
following two events can occur:
\begin{enumerate}  
\item At least one of these $\mathcal{N}_1$ individuals is detected, so the entire
  clique is isolated. The
  probability of this event is $P(D|\mathcal{N}_1)\equiv 1-(1-f)^{\mathcal{N}_1}$.
\item No individual is detected, which occurs with probability:
  \begin{eqnarray}\label{Eq.neg}
    P(\neg D|\mathcal{N}_1)=1-P(D|\mathcal{N}_1)=(1-f)^{\mathcal{N}_1}.
  \end{eqnarray}
\end{enumerate}
If the second event occurs, people from the first
generation will transmit the disease to every susceptible neighbor with
probability $\beta$. As shown in the schematic illustration (see Fig.~\ref{fig.Appshell}), these neighbors can be 
at a chemical distance of either $\ell=1$ or $\ell=2$ from the index-case. Therefore, we
split the average number of infected individuals at time $t=2$, denoted by $\langle \mathcal{N}_2 \rangle$, as:
\begin{eqnarray}
\langle \mathcal{N}_2\rangle =\epsilon_1+\epsilon_2,
\end{eqnarray}
where $\epsilon_1$ ($\epsilon_2$) corresponds to the number of new infected people at a
chemical distance of $\ell=1$ ($\ell=2$) from the index-case at time $t=2$. In what follows, we will
deduce the expressions of $\epsilon_1$ and $\epsilon_2$.

\subsubsection{Deduction of $\epsilon_2$}

\begin{figure}[H]
\vspace{0.5cm}
\begin{center}
\begin{overpic}[scale=1.5]{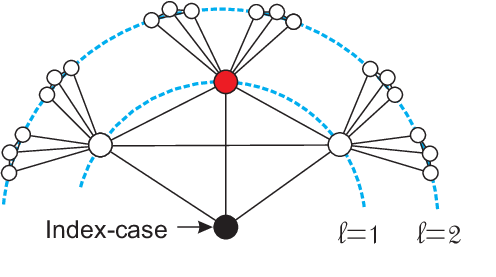}
  \put(0,65){}
\end{overpic}
\vspace{1cm}
\end{center}
\caption{(Color online) Illustration of an index-case (black node) who
  has infected one neighbor (red node) in a given clique at $t=1$. In
  this example: i) the index-case is already in a recovered state, ii)
  cliques have $k_C=4$ members, and iii) each member belongs to $k_I=3$
  cliques. Blue dashed lines indicate the chemical distance from each
  node to the index-case. Note that there are still $k_C-1-1=2$
  susceptible members (white nodes) at a chemical distance of $\ell=1$
  from the index-case.}\label{fig.Appshell}
\end{figure}

Let us assume that there are $\mathcal{N}_1$ infected people in the
first generation, and none of them have been detected. This event
occurs with probability $P(\mathcal{N}_1)P(\neg D|\mathcal{N}_1)$ (see
Eqs.~(\ref{Eq.N1}) and~(\ref{Eq.neg})). As illustrated in
Fig.~\ref{fig.Appshell}, every infected person of the first generation
has $k_I-1$ outgoing cliques, each containing $k_C-1$ susceptible
individuals. Therefore, following similar arguments leading up to
Eq.~(\ref{Eq.N1A}), we obtain that every infected person of the first
generation will transmit the disease (on average) to $(k_I-1)(k_C-1)\beta$ people
 at a distance of $\ell=2$ from the index-case.

Then, it follows that the average total number of infected individuals at a distance of
$\ell=2$ is given by,
\begin{eqnarray}
  \epsilon_2&=&\sum_{\mathcal{N}_1=0}^{k_C-1}(k_I-1)(k_C-1)\beta \mathcal{N}_1P(\mathcal{N}_1)P(\neg D|\mathcal{N}_1),\nonumber\\
  &=&(1-f)(1-\beta f)^{k_C-2}(k_I-1)(k_C-1)^2\beta^2.
\end{eqnarray}

\subsubsection{Deduction of $\epsilon_1$}

As illustrated in Fig.~\ref{fig.Appshell}, there are
$k_C-1-\mathcal{N}_1$ susceptible members at a chemical
distance of $\ell=1$ from the index-case at $t=1$. It is easy to see that, at the following
time step, the effective probability of infection for any of these members is:
\begin{eqnarray}\label{Eq.app_p}
p\equiv 1-(1-\beta)^{\mathcal{N}_1}.
\end{eqnarray}
Then, if we denote by $\mathcal{N}_2$ the number of these members who get
infected at $t=2$, we have that the probability $P(\mathcal{N}_2|\mathcal{N}_1)$ can be written as,
\begin{eqnarray}\label{Eq.N2N1}
P(\mathcal{N}_2|\mathcal{N}_1)&=&\binom{k_C-1-\mathcal{N}_1}{\mathcal{N}_2}p^{\mathcal{N}_2}(1-p)^{k_C-1-\mathcal{N}_1-\mathcal{N}_2}.
\end{eqnarray}

Finally, using Eqs.~(\ref{Eq.app_p}) and~(\ref{Eq.N2N1}), we can
estimate the average number of people becoming infected
at time $t=2$ as,
\begin{eqnarray}\label{Eq.eps1a}
\epsilon_1&=&\sum_{\mathcal{N}_1=0}^{k_C-1}\sum_{\mathcal{N}_2=0}^{k_C-1-\mathcal{N}_1}\mathcal{N}_2P(\mathcal{N}_2|\mathcal{N}_1)(1-f)^{\mathcal{N}_1}P(\mathcal{N}_1),
\end{eqnarray}
 where the factor $(1-f)^{\mathcal{N}_1}P(\mathcal{N}_1)$ is the probability
that the first generation is composed of $\mathcal{N}_1$ infected individuals and none of
them have been detected. Replacing Eqs.~(\ref{Eq.N1}) and~(\ref{Eq.N2N1}) in the last expression, and after algebraic
manipulation, we obtain,
\begin{eqnarray}
\epsilon_1=(k_C-1)(1-\beta)\left[(\beta(1-f)+(1-\beta))^{k_C-2}-(\beta(1-\beta)(1-f)+(1-\beta))^{k_C-2}\right].
\end{eqnarray}

\section{Additional results}\label{Sec.AppAddit}

In Sec.~\ref{Sec.FinS}, we investigated an SIRQ model on random networks with
cliques and showed results for $P(k_C)=\delta_{k_C,7}$ and $P(k_I)=\delta_{k_I,3}$, where
$\delta$ is the Kronecker delta. In what follows, we will study this SIRQ model for
other $P(k_C)$ and $P(k_I)$ distributions.

\subsection{RR networks with cliques}\label{Sec.appAddiRR}

Here we will present our results for RR networks with:
\begin{itemize}
\item case I)  $P(k_C)=\delta_{k_C,3}$  and $P(k_I)=\delta_{k_I,7}$,
\item case II)  $P(k_C)=\delta_{k_C,5}$   and $P(k_I)=\delta_{k_I,2}$,
\item case III)  $P(k_C)=\delta_{k_C,2}$  and $P(k_I)=\delta_{k_I,5}$.
\end{itemize}
For the simulations, only one person is infected at the beginning of
the dynamic process.

\begin{figure}[H]
\vspace{0.5cm}
\begin{center}
\begin{overpic}[scale=0.25]{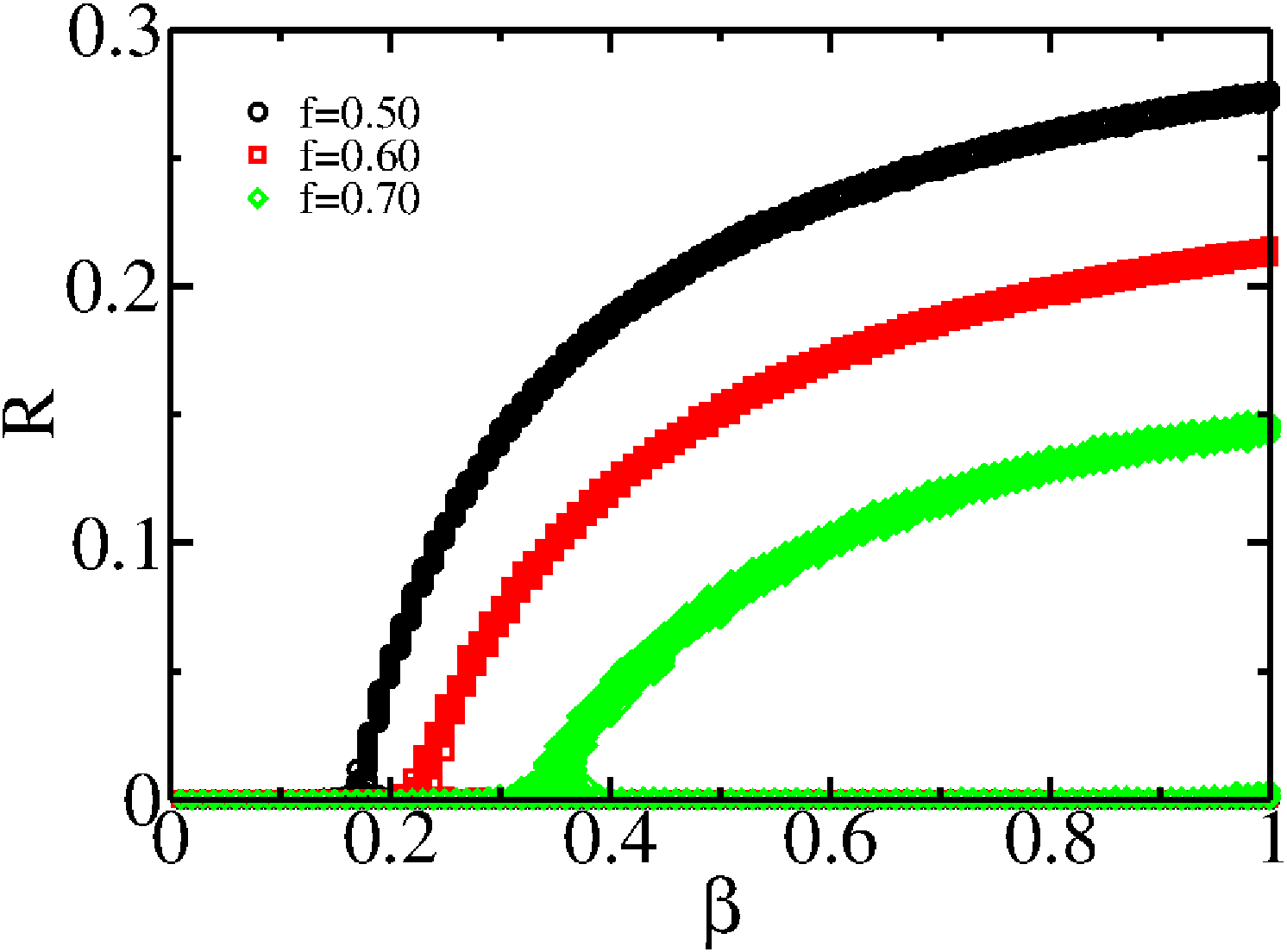}
  \put(20,20){(a)}
\end{overpic}
\vspace{0.5cm}
\begin{overpic}[scale=0.25]{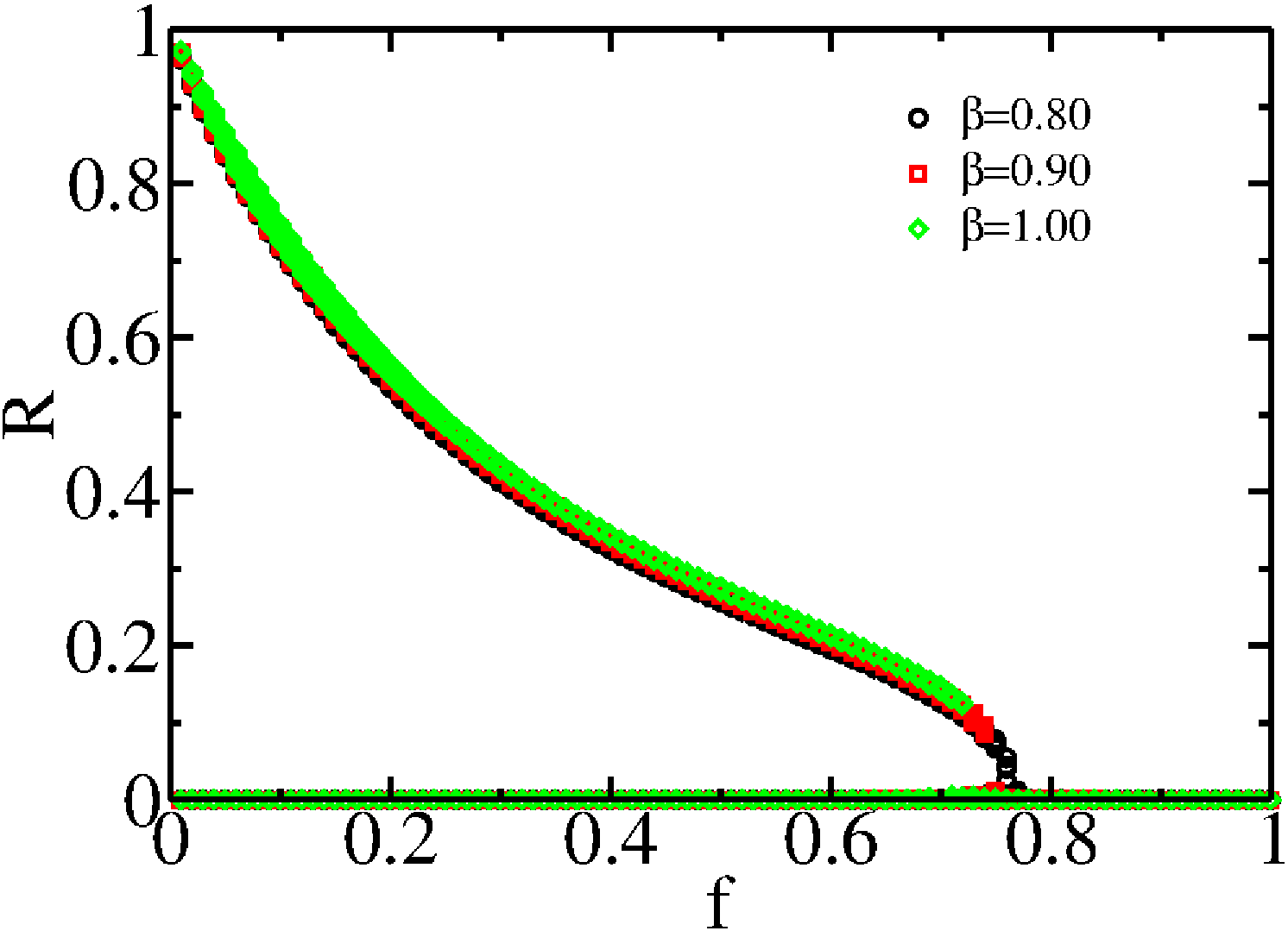}
  \put(80,20){(b)}
\end{overpic}
\vspace{0.0cm}
\begin{overpic}[scale=0.25]{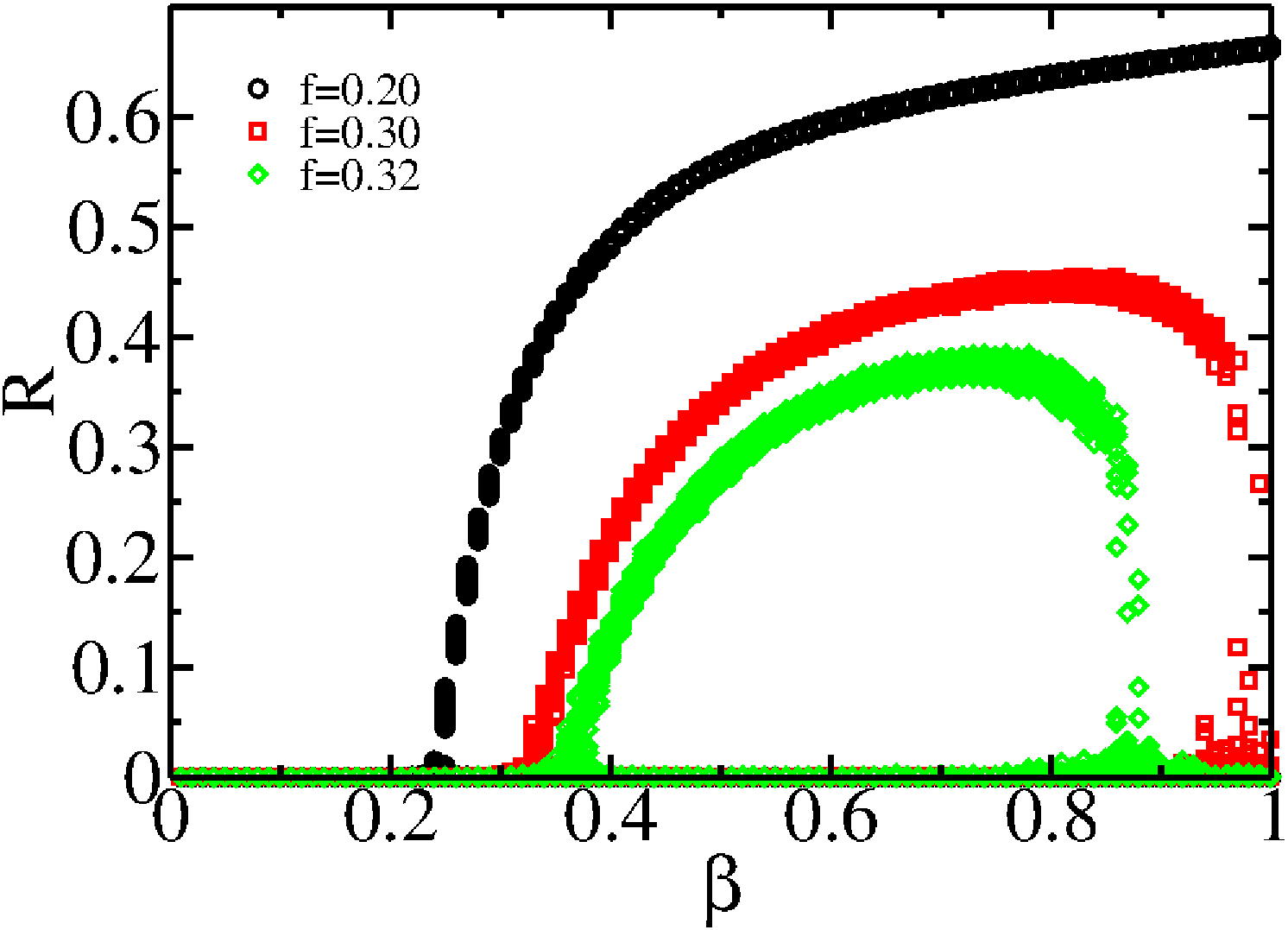}
  \put(20,20){(c)}
\end{overpic}
\begin{overpic}[scale=0.25]{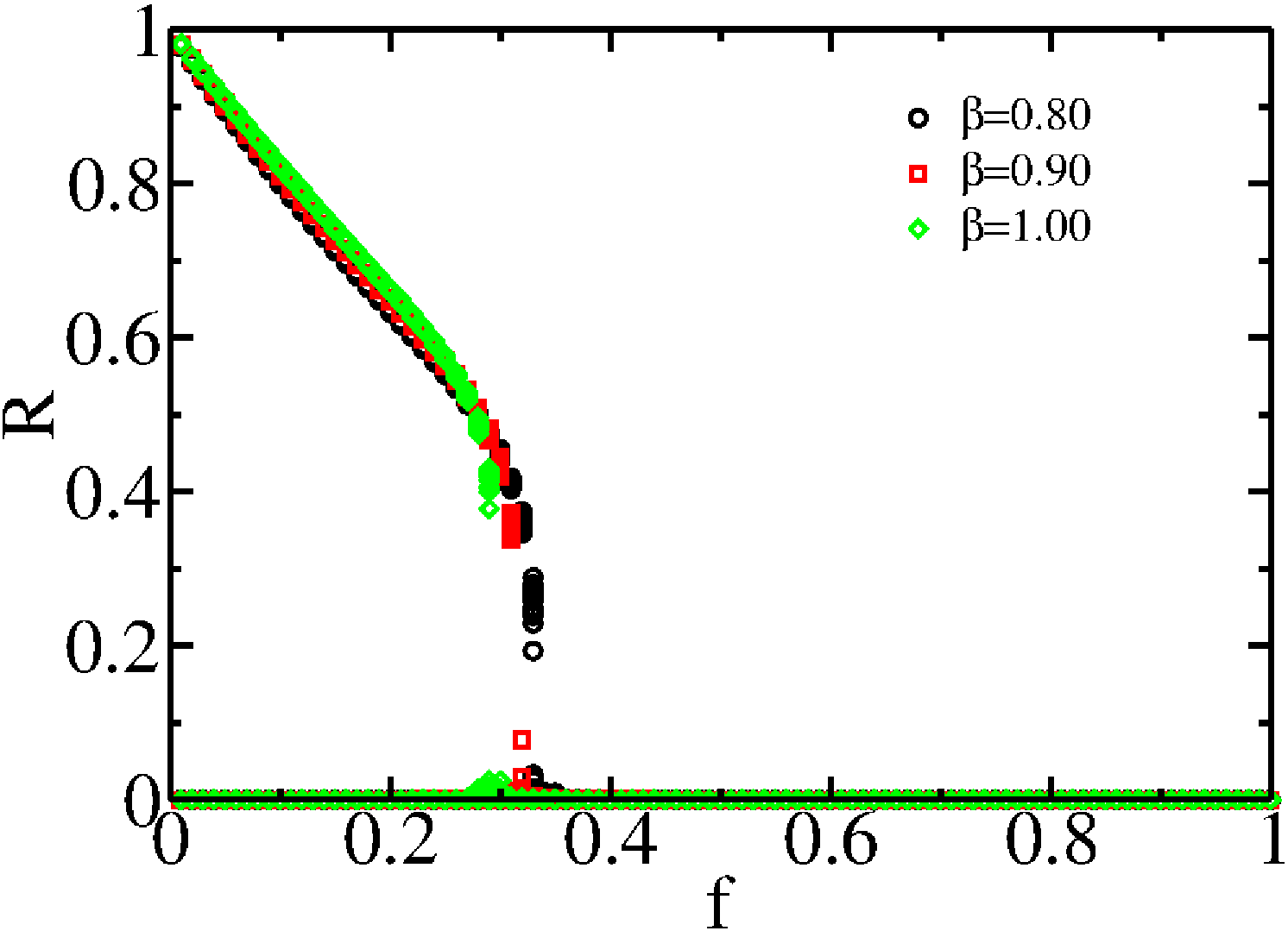}
  \put(80,20){(d)}
\end{overpic}
\vspace{1cm}
\begin{overpic}[scale=0.25]{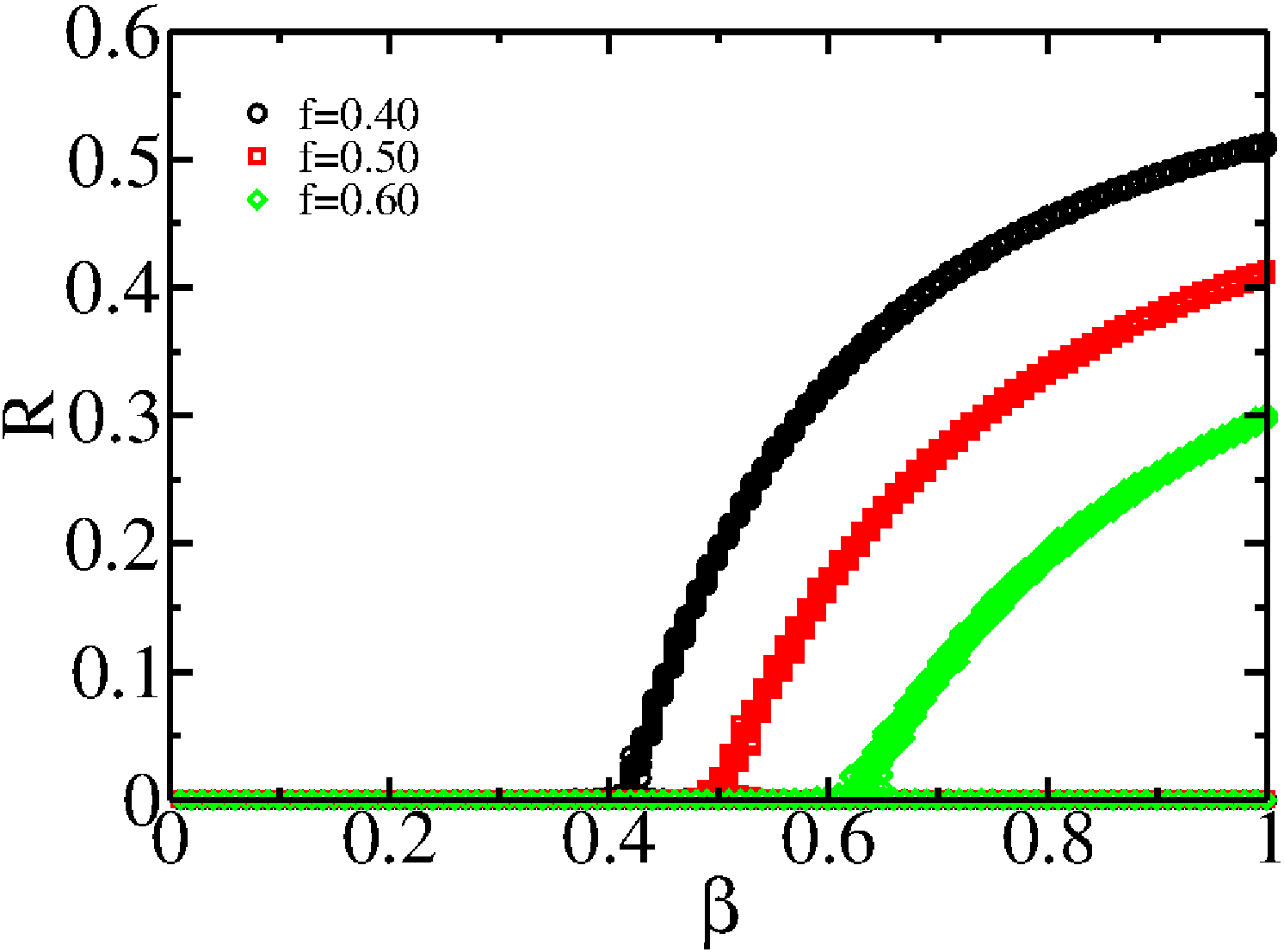}
  \put(20,20){(e)}
\end{overpic}
\begin{overpic}[scale=0.25]{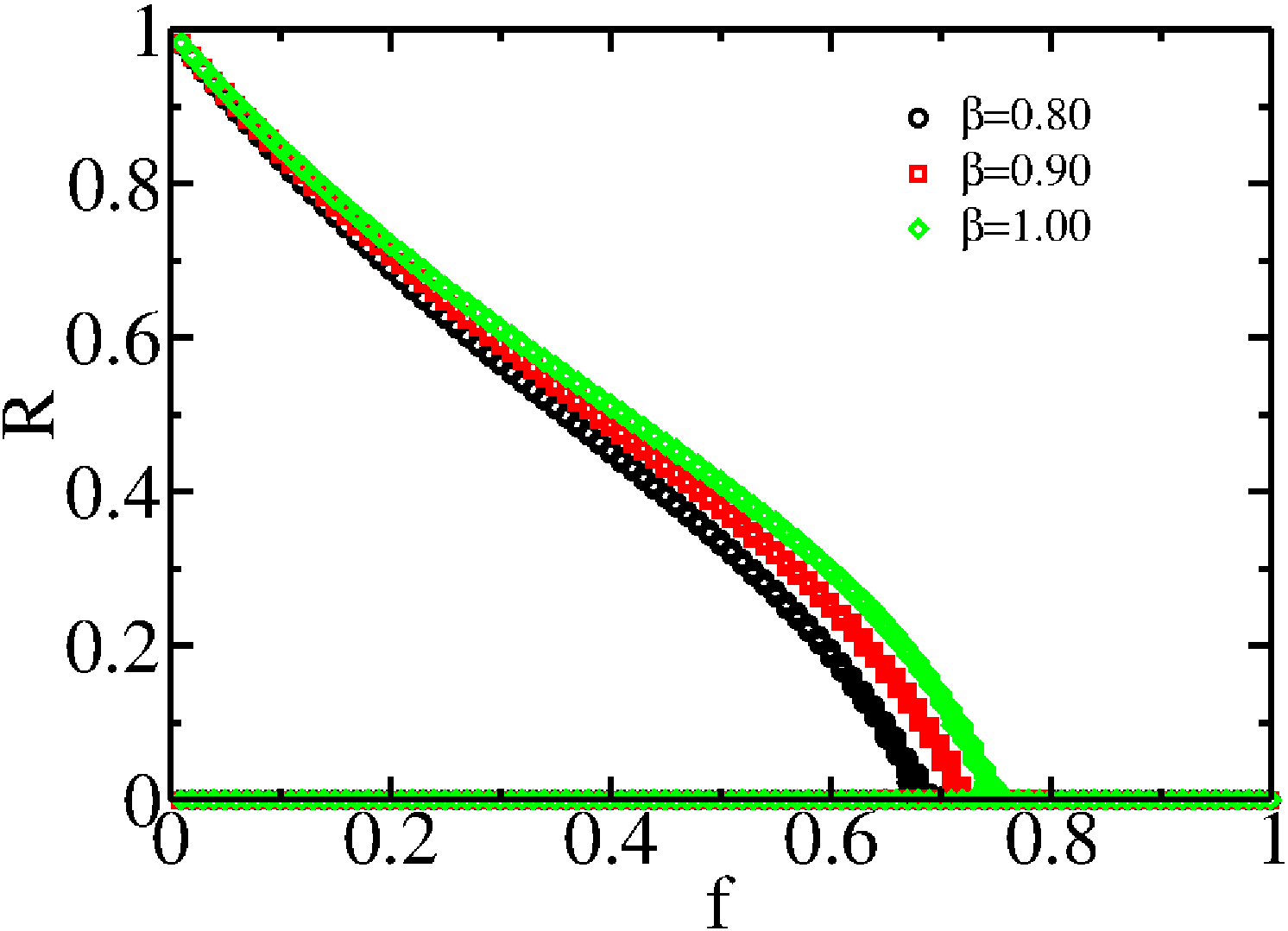}
  \put(80,20){(f)}
\end{overpic}
\vspace{1cm}
\vspace{0.0cm}
\end{center}
\caption{(Color online) Scatter plot of the fraction of recovered
  population $R$ at the final stage as a function of $\beta$ (panels
  a, c, e), and as a function of $f$ (panels b, d, f). Results were
  obtained from $10^3$ stochastic realizations on RR networks with:
  $k_C=3$ and $k_I=7$ (panels a and b), $k_C=5$, $k_I=2$ (panels c and
  d), and $k_C=2$, $k_I=5$ (panels e and f). }\label{fig.appScatRR}
\end{figure}

In Figs.~\ref{fig.appScatRR}a,c,e, we show a scatter-plot of the fraction of recovered
people $R$ at the final stage as a function of $\beta$ for
several values of $f$. Similarly, in Figs.~\ref{fig.appScatRR}b,d,f, we show $R$ at the
final stage as a function of $f$ for different values of $\beta$.

In contrast to what was observed in Sec.~\ref{Sec.FinS}, here we do
not find any abrupt transition for case III, as shown in
Figs.~\ref{fig.appScatRR}e,f . This result may be due to the fact
that cliques are smaller in case III than in cases I and II,
so these cliques have a lower probability of being quarantined. On the other hand, for cases I and II, we obtain
that $R$ exhibits an abrupt transition for high values of $\beta$, as
seen in Figs.~\ref{fig.appScatRR}b,d.

\begin{figure}[H]
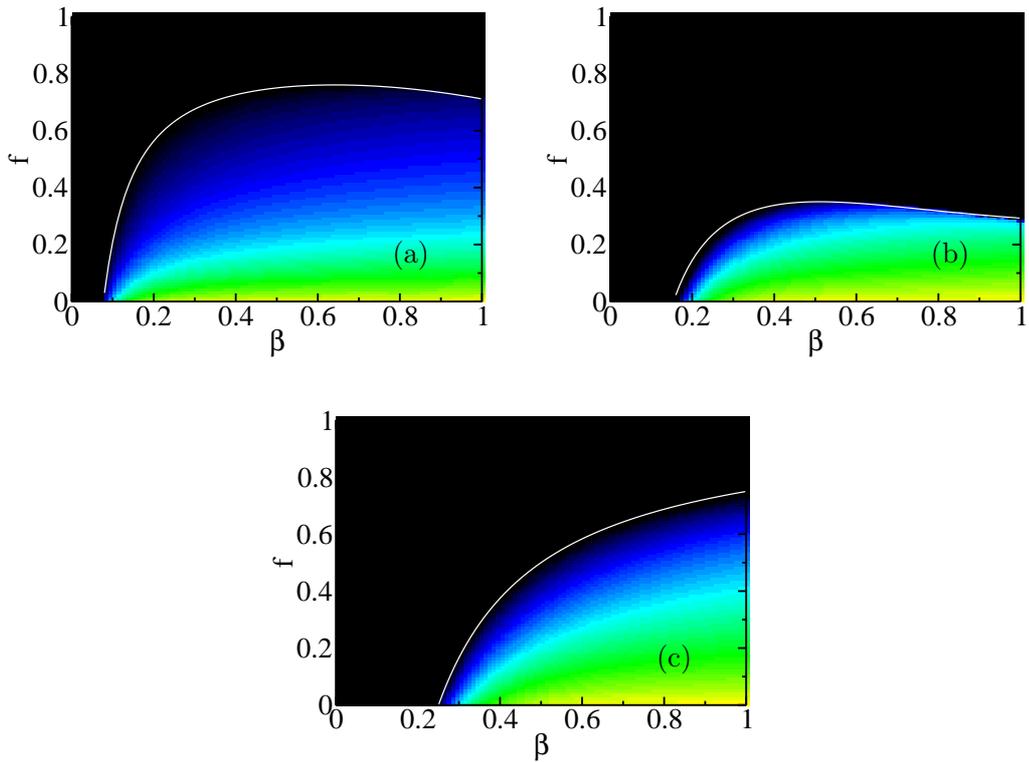

\vspace{0.5cm}
\begin{center}
\begin{overpic}[scale=0.25]{Fig09a.eps}
  \put(80,20){(a)}
\end{overpic}
\hspace{0.5cm}
\vspace{0.7cm}
\begin{overpic}[scale=0.25]{Fig09b.eps}
  \put(80,20){(b)}
\end{overpic}
\vspace{0.5cm}
\begin{overpic}[scale=0.25]{Fig09c.eps}
  \put(80,20){(c)}
\end{overpic}
\vspace{0.5cm}
\end{center}
\caption{(Color online) Heat-map of $R$ in the plane $\beta-f$ for a RR with cliques with: $k_C=3$ and $k_I=7$
  (panel a), $k_C=5$ and $k_I=2$ (panel b), and  $k_C=2$ and $k_I=5$ (panel c), obtained from stochastic
  simulations for networks with $N_I=10^6$. Simulation results were averaged over $10^3$ stochastic
  realizations. The solid white line was obtained from
  Eq.~(\ref{eq.R0}) for $R_0=1$.}\label{fig.appPhase}
\end{figure}

In Figs.~\ref{fig.appPhase}a-c, we plot the heat-map of the fraction
of recovered individuals when an epidemic occurs ($R>1\%$) in the
plane $\beta-f$. In these figures, we also include the curve where
$R_0=1$ [obtained from Eq.~(\ref{eq.R0})]. For all cases, we observe
that this curve predicts well the boundary between the epidemic and
non-epidemic phases.

\begin{figure}[H]
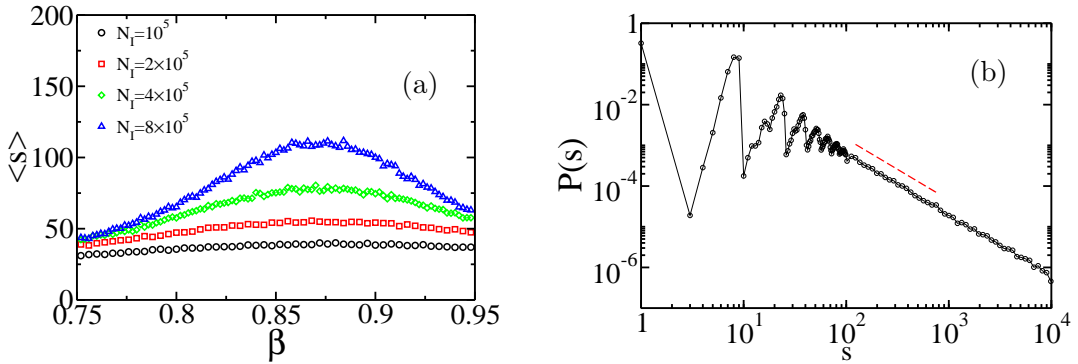

\vspace{0.5cm}
\begin{center}
\begin{overpic}[scale=0.25]{Fig10a.eps}
  \put(80,55){(a)}
\end{overpic}
\vspace{0.5cm}
\hspace{0.5cm}
\begin{overpic}[scale=0.25]{Fig10b.eps}
  \put(80,55){(b)}
\end{overpic}
\vspace{0.5cm}
\vspace{0.0cm}
\end{center}
\caption{(Color online) Panel a: $\langle s\rangle$ against $\beta$
  (fixing $f=0.32$) for RR networks with cliques with $k_C=5$, $k_I=2$ and several
  values of $N_I$. Results were averaged over $10^5$
  realizations. Panel b: distribution $P(s)$ for $\beta=0.87$,
  $f=0.32$, and $N_I=10^6$, obtained from $1.5\times 10^5$ stochastic
  realizations (symbols). The solid black line is a guide to the eye,
  and the dashed red line is a power-law function with exponent
  $\tau-1=1.5$.}\label{fig.appOutRR52}
\end{figure}

Now for case II, we measure:
\begin{enumerate}
\item the average size of small outbreaks
$\langle s\rangle$ vs. $\beta$ for $f=0.32$ and different
  network sizes $N_I$ (see Fig.~\ref{fig.appOutRR52}a),
\item the distribution of small outbreak sizes
  for $\beta=0.87$ and $f=0.32$ (see Fig.~\ref{fig.appOutRR52}b).
\end{enumerate}  
As in Figs.~\ref{fig.Pinf}b-c, we can see that $\langle s \rangle$ has
a peak around $\beta=0.87$, and $P(s)$ decays as a
power-law. Therefore, our findings reveal that outbreaks exhibit features of a continuous phase transition around $\beta=0.87$ for case
II.

\subsection{Non-regular random networks with cliques}

So far, we have focused our attention on random networks with cliques in
which $k_C$ and $k_I$ follow a delta distribution. In this section, we
show results for random networks with cliques in which $k_C$ and $k_I$
follow other probability distributions. Specifically, we consider
\begin{itemize}
\item a truncated Poisson distribution, defined as
\begin{equation}
Pois(\lambda,k_{min},k_{max}) =
    \begin{cases}
        c\frac{\lambda^ke^{-\lambda}}{k!} & \text{if } k_{min}\leq k\leq k_{max}\\
        0 & \text{otherwise,}
    \end{cases}
\end{equation}
where $c$ is a normalization constant;
\item a truncated power-law distribution, defined as
\begin{equation}
PL(\lambda,k_{min},k_{max}) =
    \begin{cases}
        c k^{-\lambda} & \text{if } k_{min}\leq k\leq k_{max}\\
        0 & \text{otherwise,}
    \end{cases}
\end{equation}
where $c$ is a normalization constant.
\end{itemize}

\begin{figure}[H]
\vspace{0.5cm}
\begin{center}
\begin{overpic}[scale=0.25]{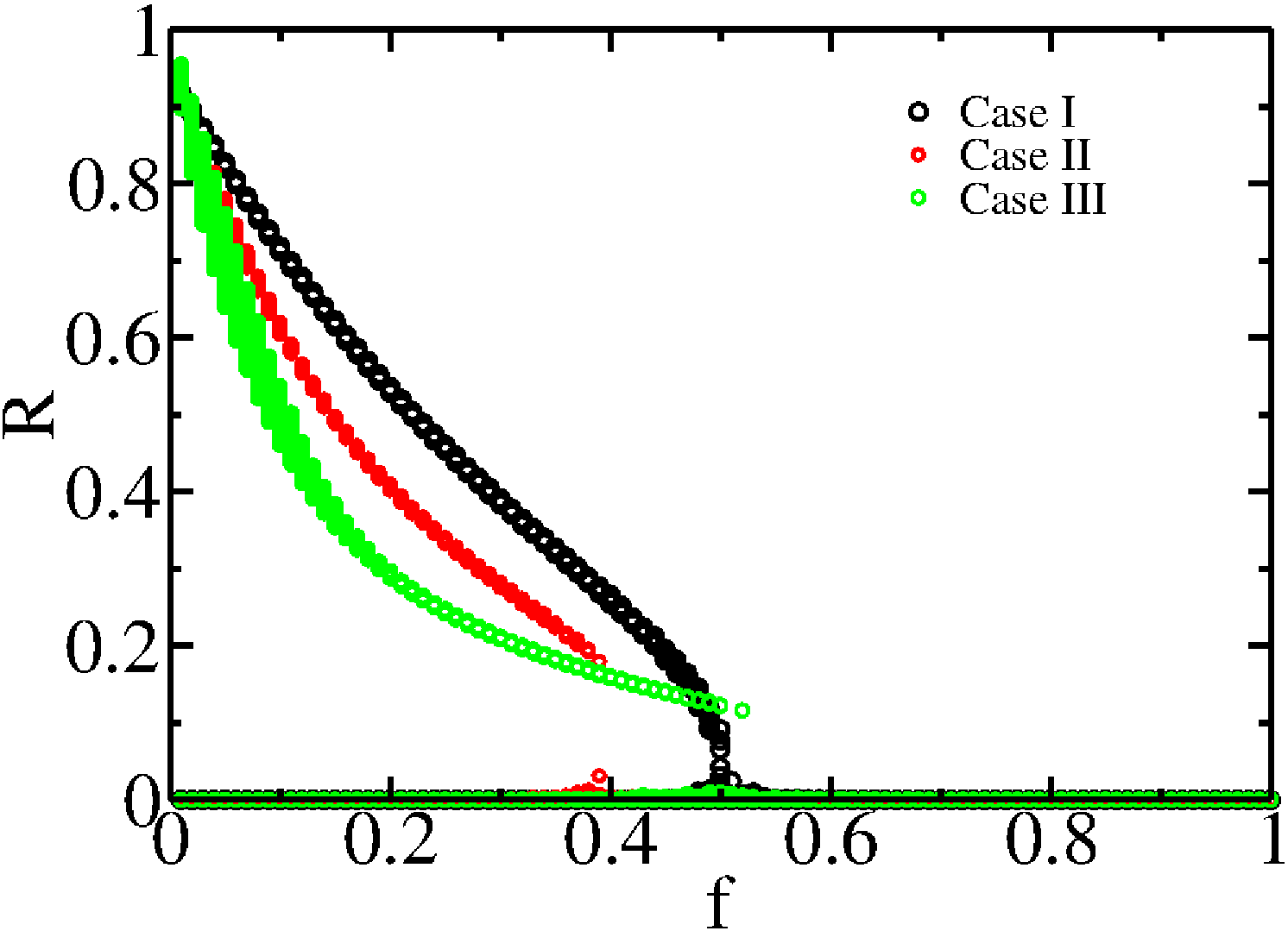}
  \put(80,20){(a)}
\end{overpic}
\vspace{0.5cm}
\hspace{0.5cm}
\begin{overpic}[scale=0.25]{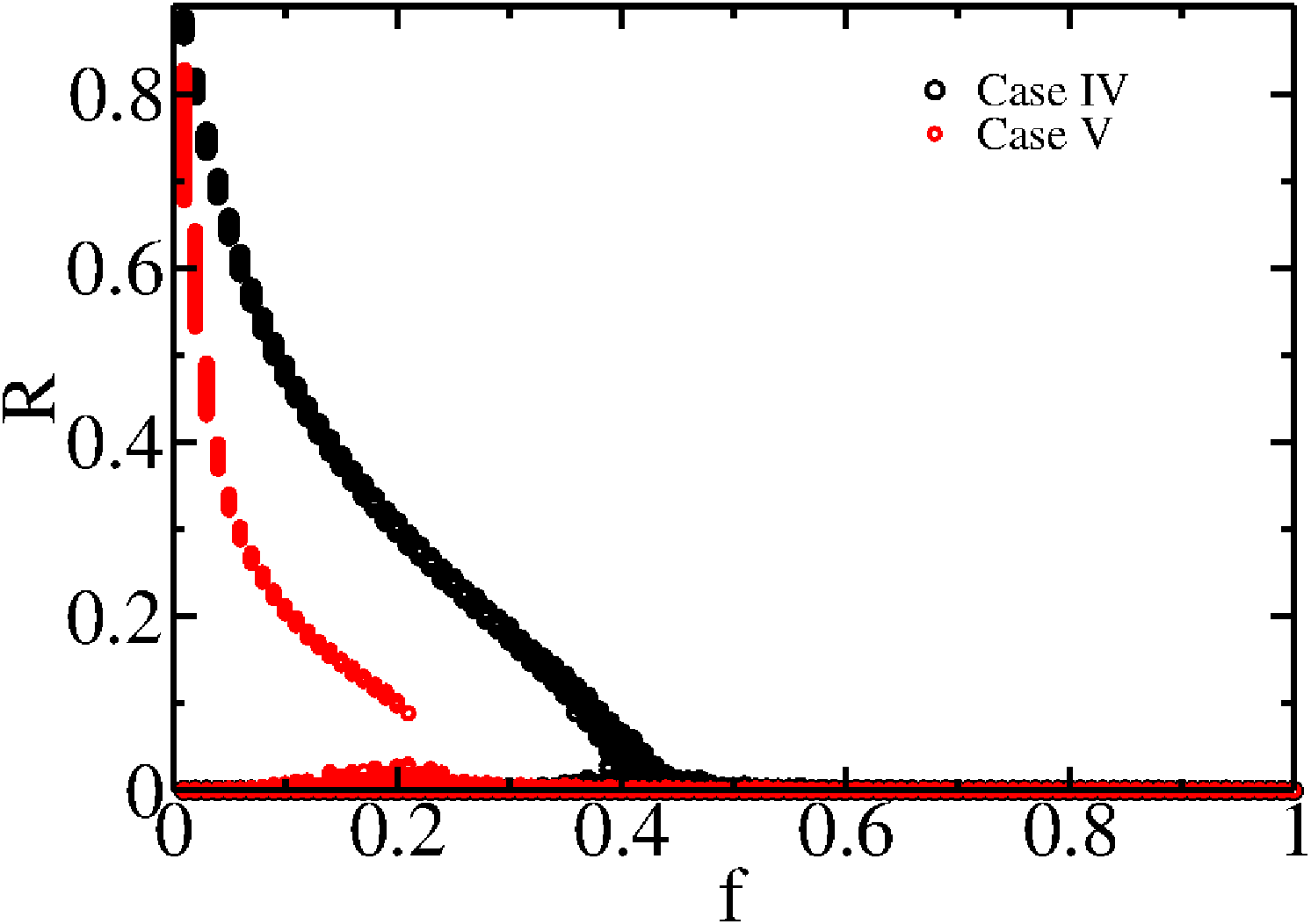}
  \put(80,20){(b)}
\end{overpic}
\vspace{1cm}
\vspace{0.0cm}
\end{center}
\caption{(Color online) Scatter plot of the fraction of recovered
  population $R$ at the final stage as a function of $f$ for: cases
  I-III (panel a), and cases IV and V (panel b) --see Table~\ref{app.TableCases}. Results were obtained from $10^3$ stochastic realizations on
  random networks with $N_I=10^6$. }\label{fig.appScatnonRR}
\end{figure}

In Figs.~\ref{fig.appScatnonRR}a-b, we plot the fraction of recovered
people at the final stage vs. $f$ (fixing $\beta=1$) for the cases
listed in Table~\ref{app.TableCases}. In this table, we also include
the mean value of $k_C$ (i.e., $\langle
k_C\rangle=\sum_{k_C}k_CP(k_C)$) and its variance ($VAR(k_C)=\langle
k_C^2\rangle-\langle k_C\rangle^2$):
\begin{table}[H]
\caption{Cases considered in Figs.~\ref{fig.appScatnonRR}a-b.}
\label{app.TableCases}
\begin{tabular}{l|l|l|l|l}
Case & $P(k_C)$ & $P(k_I)$ & $\langle k_C\rangle$ & $VAR(k_C)$  \\
I  & $Pois(3,0,20)$ & $Pois(3,0,20)$  & 3 & 3 \\
II  & $Pois(7,0,20)$   & $Pois(3,0,20)$  & 7 & 7 \\
III  & $Pois(7,0,20)$  & $Pois(7,0,20)$  & 7 & 7\\
IV  & $PL(2.0,2,100)$ & $Pois(3,0,20)$ & 6.4 & 112.4\\
V  & $PL(1.5,2,100)$ & $Pois(3,0,20)$  & 12.4 & 319.5
\end{tabular}
\end{table}

Similarly to the results in Sec.~\ref{Sec.appAddiRR}, from
Figs.~\ref{fig.appScatnonRR}a-b we note that as the average clique
size increases, the system tends to exhibit an abrupt transition.

\begin{figure}[H]
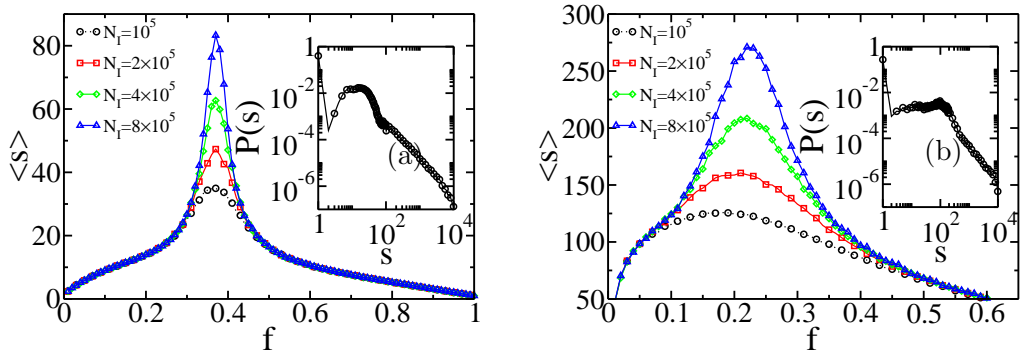

\vspace{0.5cm}
\begin{center}
\begin{overpic}[scale=0.25]{Fig12a.eps}
  \put(80,40){(a)}
\end{overpic}
\vspace{0.5cm}
\hspace{0.5cm}
\begin{overpic}[scale=0.25]{Fig12b.eps}
  \put(80,40){(b)}
\end{overpic}
\vspace{0.0cm}
\end{center}
\caption{(Color online) $\langle s\rangle$ against $f$ (fixing $\beta=1$) for several
  values of $N_I$. Panels a and b correspond to cases II and V,
  respectively.  Results were averaged over $10^4$ realizations. In
  the insets, we show the distribution $P(s)$ (in log-log scale) for: $\beta=1$, $f=0.36$,
  and $N_I=10^6$ (panel a), and for $\beta=1$, $f=0.22$, and
  $N_I=10^6$ (panel b). These results were averaged over $2\times 10^5$ and $4\times 10^4$ stochastic
  realizations for panels a and b, respectively. Solid lines are a guide to the
  eye.}\label{fig.appSmenns_nonRR}
\end{figure}

Now, focusing on cases II and V, in Fig.~\ref{fig.appSmenns_nonRR}, we
plot $\langle s \rangle$ vs. $f$ (fixing
$\beta=1$). Additionally, in the insets we show the probability
distribution of the final number of recovered people for:
\begin{itemize}
\item $\beta=1$ and $f=0.36$ (for case II),
\item $\beta=1$ and $f=0.22$ (for case V).
\end{itemize}
As in Secs.\ref{Sec.FinS} and~\ref{Sec.appAddiRR}, we obtain that $\langle s\rangle$ has a peak around $f_c(N_I)$ and $P(s)$ decays as a power-law, so these results indicate that
outbreaks exhibit some features of a continuous phase transition.

In summary, our findings suggest that non-regular networks with
larger cliques tend to exhibit features of both discontinuous and
continuous phase transitions.

\section{Probability of a small outbreak}\label{Sec.AppPi}

In Sec.~\ref{sec.prob}, we calculated the probability of a small
outbreak, $1-\Pi$, as a function of $f$ (with $\beta=1$) for the case in which
$k_C$ and $k_I$ follow a delta distribution. Here, we will
generalize our previous equations to compute $1-\Pi$ for any arbitrary probability
distributions $P(k_C)$ and $P(k_I)$.

Let us consider a random network with cliques that can be represented
by a bipartite network. We define the following generating functions:
\begin{enumerate}
\item $G_{0C}[x]=\sum_{k_C}P(k_C)x^{k_C}$, which is the generating
function for the distribution $P(k_C)$, and $G_{1C}[x]=\sum_{k_C}k_C
P(k_C)/\langle k_C\rangle x^{k_C-1}$ which is the generating function
for the distribution $k_C P(k_C)/\langle k_C\rangle$,
\item $G_{0I}[x]=\sum_{k_I}P(k_I)x^{k_I}$, which is the generating
  function for the distribution $P(k_I)$, and $G_{1I}[x]=\sum_{k_I}k_I
  P(k_I)/\langle k_I\rangle x^{k_I-1}$ which is the generating
  function for the distribution $k_I P(k_I)/\langle
  k_I\rangle$.
\end{enumerate}

In the same way as in Sec.~\ref{sec.prob}, we define $\phi$ as the
probability that an infected individual (reached by a link) does not
generate an epidemic. For any arbitrary distribution $P(k_C)$ and $P(k_I)$, we
have that $\phi$ obeys the following equation:
\begin{eqnarray}\label{eq.Appphi}
  \phi&=&G_{1I}\left[ G_{1C}[(1-f)\phi]+1-G_{1C}[1-f]\right],\\
  &=&\sum_{k_I}\frac{k_I  P(k_I)}{\langle k_I\rangle} \left( G_{1C}[(1-f)\phi]+1-G_{1C}[1-f]   \right)^{k_I-1},
\end{eqnarray}
The right-hand side of the above equation is the probability that
an individual ``$j$'' (reached through a link) does not generate an
epidemic. This is due to the fact that none of the $k_I-1$ outgoing cliques
generates an epidemic, because one of the following two events occurs in every clique:
\begin{enumerate}
\item the outgoing clique has at least one infected member who is detected  with probability $1-G_{1C}[1-f]$, so the entire clique is quarantined.
\item the outgoing clique is not isolated, but it will not be able to cause
  an epidemic with probability $G_{1C}[(1-f)\phi]$.
\end{enumerate}
On the other hand, the probability $1-\Pi$ that a randomly chosen individual will
not generate an epidemic is given by:
\begin{eqnarray}\label{eq.AppnoPi}
  1-\Pi=f+(1-f)G_{0I}\left[ G_{1C}[(1-f)\phi]+1-G_{1C}[1-f]\right],
\end{eqnarray}
where the second term on the right-hand side of this equation has an
  interpretation similar to that of
Eq.~(\ref{eq.Appphi}). Following a procedure similar to that in
Sec.~\ref{sec.prob}, it can be seen that the critical probability of
detection $f_c$ at which a phase transition occurs is implicitly given by
\begin{eqnarray}\label{eq.appfc}
1 = (1-f_c) G_{1I}^{'}(1)G_{1C}^{'}[(1-f_c)],
\end{eqnarray}
where $G_{1I}^{'}(x)\equiv dG_{1I}/dx$, and $G_{1C}^{'}(x)\equiv
dG_{1C}/dx$.  In particular, for the case where $k_C$ and $k_I$ follow
a Poisson distribution (i.e., $Pois(\lambda,0,\infty)$), the above
equation can be solved explicitly in terms of the Lambert
W-function~\footnote{Here, we use the principal branch of the Lambert
function.}:
\begin{equation}
    f_c = 1-\frac{W\left(\frac{e^{\langle k_C\rangle}}{\langle k_I\rangle}\right)}{\langle k_C\rangle}.
\end{equation}

\begin{figure}[H]
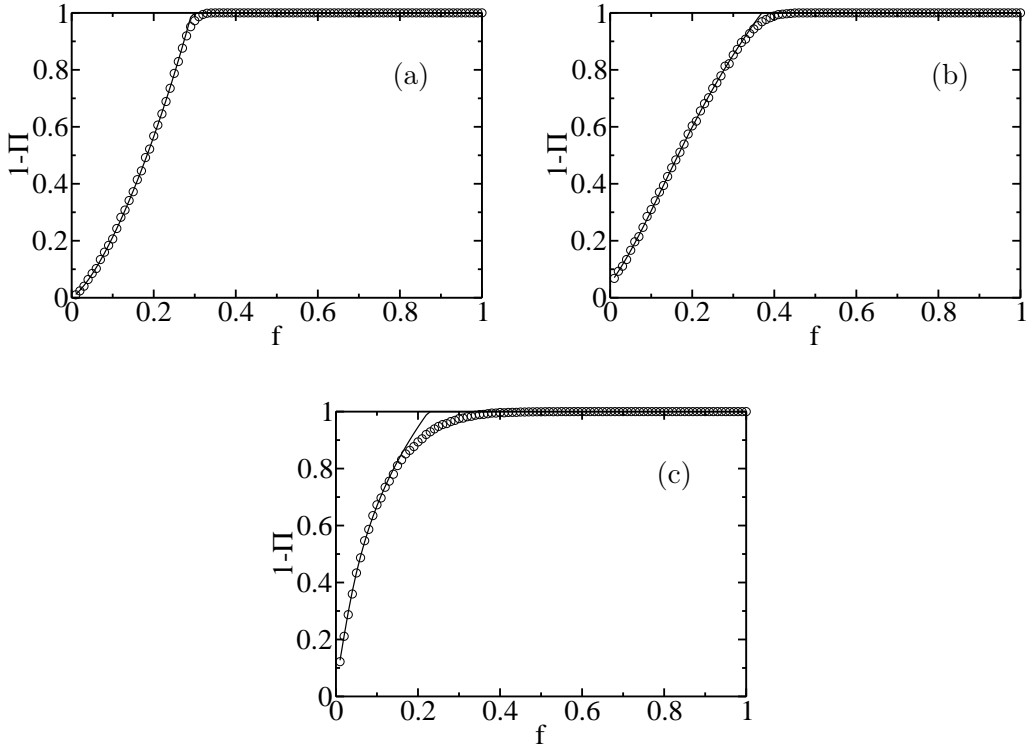

\vspace{0.5cm}
\begin{center}
\begin{overpic}[scale=0.25]{Fig13a.eps}
  \put(80,55){(a)}
\end{overpic}
\vspace{0.7cm}
\hspace{0.5cm}
\begin{overpic}[scale=0.25]{Fig13b.eps}
  \put(80,55){(b)}
\end{overpic}
\vspace{0.5cm}
\begin{overpic}[scale=0.25]{Fig13c.eps}
  \put(80,55){(c)}
\end{overpic}
\vspace{0.5cm}
\end{center}
\caption{Probability of a small outbreak ($1-\Pi$) vs.
  $f$ for $\beta=1$ and several network topologies: RR networks
  with $P(k_C)=\delta_{k_C,5}$ and $P(k_I)=\delta_{k_I,2}$ (panel a),
  non-RR networks with $P(k_C)=Pois(7,0,20)$ and $P(k_I)=Pois(3,0,20)$
  (panel b), and non-RR networks with $P(k_C)=PL(1.5,2,100)$ and
  $P(k_I)=Pois(3,0,20)$ (panel c). The line corresponds to the theory
  given by Eqs.~(\ref{eq.Appphi})-(\ref{eq.AppnoPi}), and symbols are
  simulation results averaged over $10^4$ network realizations with
  $N_I=10^5$.}\label{fig.appPi}
\end{figure}

To see the validity of our equations, we run numerical simulations
for the following networks:
\begin{itemize}
\item case I: RR networks with $P(k_C)=\delta_{k_C,5}$   and $P(k_I)=\delta_{k_I,2}$, where $f_c=0.293$ (computed from  Eq.~(\ref{eq.appfc})),
\item case II: non-RR networks with $P(k_C)=Pois(7,0,20)$   and $P(k_I)=Pois(3,0,20)$, where $f_c=0.369$,
\item case III: non-RR networks with $P(k_C)=PL(1.5,2,100)$   and $P(k_I)=Pois(3,0,20)$, where $f_c=0.226$. 
\end{itemize}
In Fig.~\ref{fig.appPi}, we show the probability of an outbreak as a
function of $f$ (with $\beta=1$) obtained from our simulations and
Eqs.~(\ref{eq.Appphi})-(\ref{eq.AppnoPi}). As we can see, the
agreement between theory and simulations is excellent.

\section{Backward bifurcation: additional results}\label{sec.appBif}
 In Sec.~\ref{Sec.back}, we found (for $\beta=1$) that a backward
 bifurcation phenomenon emerges for RR networks with cliques with
 $k_C=7$ and $k_I=3$. Here, we will show that this phenomenon also
 occurs on networks where $k_C$ and $k_I$ follow other probability
 distributions.

\begin{figure}[H]
\vspace{0.5cm}
\begin{center}
\begin{overpic}[scale=0.25]{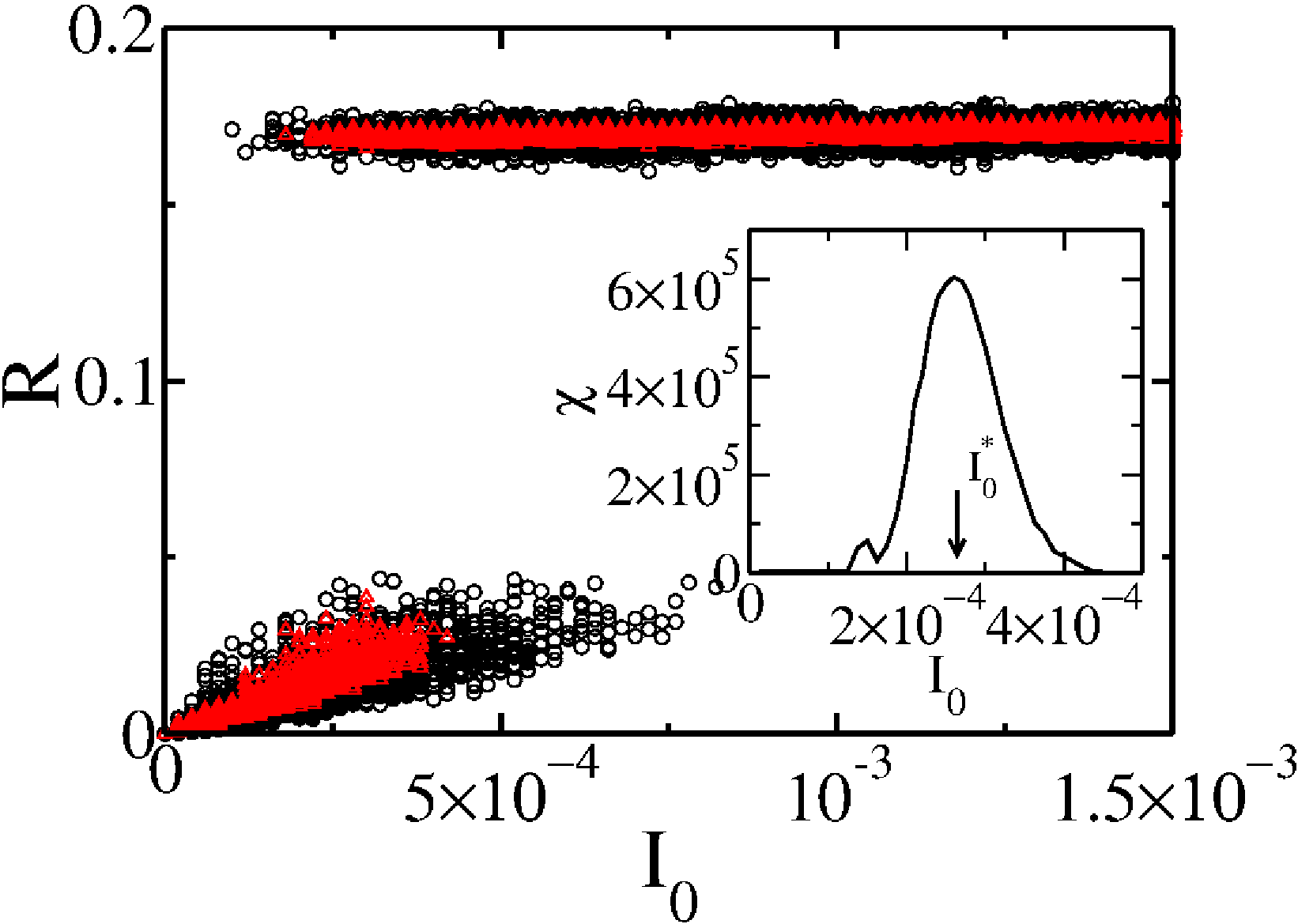}
  \put(80,48){(a)}
\end{overpic}
\vspace{0.5cm}
\hspace{0.5cm}
\begin{overpic}[scale=0.25]{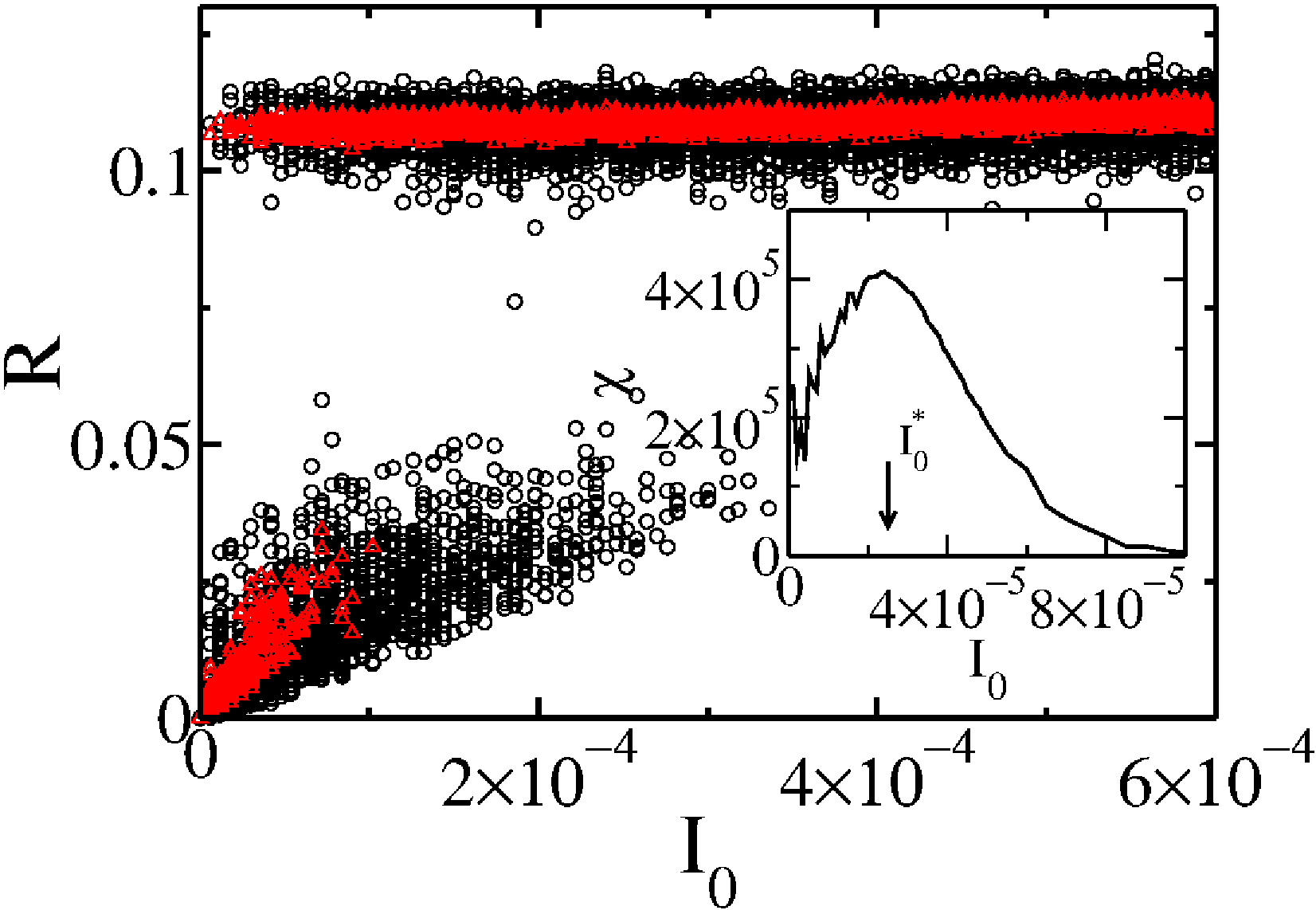}
  \put(80,48){(b)}
\end{overpic}
\vspace{0.5cm}
\end{center}
\caption{(Color online) Scatter-plot of $R$ vs. $I_0$ (fixing
  $\beta=1$) obtained from numerical simulations on random networks
  where: $P(k_C)=Pois(7,0,20)$ and $P(k_I)=Pois(3,0,20)$ (panel a),
  and $P(k_C)=PL(1.5,2,100)$ and $P(k_I)=Pois(3,0,20)$ (panel b). We
  set the probability of detection to $f=0.40$ for panel a, and
  $f=0.25$ for panel b. Note that in both cases, we used $f>f_c$ (see
  the value of $f_c$ in Appendix~\ref{Sec.AppPi}). Numerical results
  were obtained from $10^3$ stochastic realizations  (panel a) and $3\times 10^3$ stochastic realizations (panel b) on networks with
  $N_I=10^6$ (black symbols) and $N_I=8\times 10^6$ (red symbols). The insets show the susceptibility of the number of recovered people, $\chi$, as a function of $I_0$ (for $N_I=8\times 10^6$), where $I_0^*$ is the peak position.  }\label{fig.appBacki}
\end{figure}

 In Fig.~\ref{fig.appBacki}, we plot the fraction of recovered individuals at the
 final stage as a function of $I_0$ (fixing $\beta=1$) for the
 following cases:
\begin{enumerate}
\item case I: $P(k_C)=Pois(7,0,20)$ and $P(k_I)=Pois(3,0,20)$,
\item case II: $P(k_C)=PL(1.5,2,100)$ and $P(k_I)=Pois(3,0,20)$.
\end{enumerate}

As in Sec.~\ref{Sec.back}, we can see that $R$ exhibits an abrupt jump
around a threshold $I_0^*$ for $\beta=1$, where $I_0^*$ is the peak position of the susceptibility~\cite{ferreira2012epidemic}, defined as
\begin{eqnarray}
    \chi=N_I\frac{\langle R^2\rangle-\langle R\rangle^2}{\langle R\rangle}.
\end{eqnarray}
Similar results can be obtained for
$\beta=0.8$ and $\beta=0.90$ (not shown here).

\begin{figure}[H]
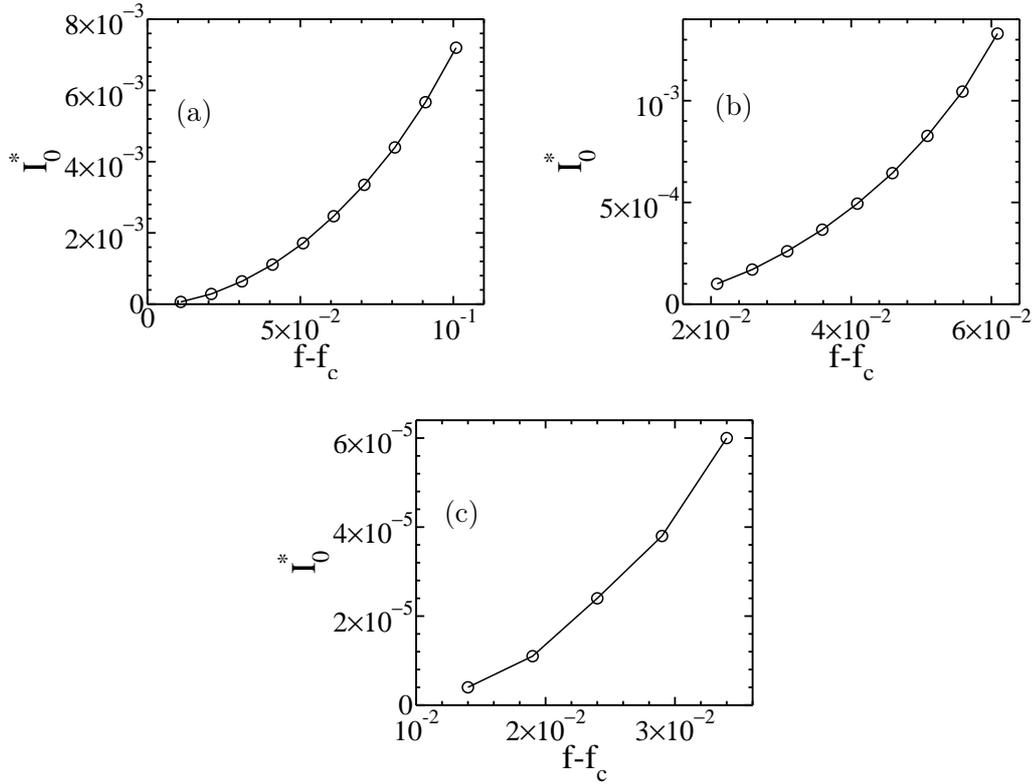

\vspace{0.5cm}
\begin{center}
\begin{overpic}[scale=0.25]{Fig15a.eps}
  \put(35,55){(a)}
\end{overpic}
\vspace{0.5cm}
\hspace{0.5cm}
\begin{overpic}[scale=0.25]{Fig15b.eps}
  \put(35,55){(b)}
\end{overpic}
\vspace{0.5cm}
\begin{overpic}[scale=0.25]{Fig15c.eps}
  \put(35,55){(c)}
\end{overpic}
\vspace{0.5cm}
\end{center}
\caption{$I_0^*$ as a function of $f-f_c$ (fixing $\beta=1$) for
  random networks with cliques where: $P(k_C)=\delta_{k_C,7}$ and
  $P(k_I)=\delta_{k_I,3}$ (panel a), $P(k_C)=Pois(7,0,20)$ and
  $P(k_I)=Pois(3,0,20)$ (panel b), and $P(k_C)=PL(1.5,2,100)$ and
  $P(k_I)=Pois(3,0,20)$ (panel c).  Symbols represent simulation
  results in networks with $N_I=8\times 10^6$ and solid lines are a guide to the eye. The number of realizations was:  250 (panel a), 1000 (panel b),  and 3000 (panel c).}\label{fig.appBackI0}
\end{figure}

Finally, in Fig.~\ref{fig.appBackI0}, we display the threshold $I_0^*$
vs. $f-f_c$ for different network structures.

\bibliography{bib}

\begin{thebibliography}{56}%
\makeatletter
\providecommand \@ifxundefined [1]{%
 \@ifx{#1\undefined}
}%
\providecommand \@ifnum [1]{%
 \ifnum #1\expandafter \@firstoftwo
 \else \expandafter \@secondoftwo
 \fi
}%
\providecommand \@ifx [1]{%
 \ifx #1\expandafter \@firstoftwo
 \else \expandafter \@secondoftwo
 \fi
}%
\providecommand \natexlab [1]{#1}%
\providecommand \enquote  [1]{``#1''}%
\providecommand \bibnamefont  [1]{#1}%
\providecommand \bibfnamefont [1]{#1}%
\providecommand \citenamefont [1]{#1}%
\providecommand \href@noop [0]{\@secondoftwo}%
\providecommand \href [0]{\begingroup \@sanitize@url \@href}%
\providecommand \@href[1]{\@@startlink{#1}\@@href}%
\providecommand \@@href[1]{\endgroup#1\@@endlink}%
\providecommand \@sanitize@url [0]{\catcode `\\12\catcode `\$12\catcode
  `\&12\catcode `\#12\catcode `\^12\catcode `\_12\catcode `\%12\relax}%
\providecommand \@@startlink[1]{}%
\providecommand \@@endlink[0]{}%
\providecommand \url  [0]{\begingroup\@sanitize@url \@url }%
\providecommand \@url [1]{\endgroup\@href {#1}{\urlprefix }}%
\providecommand \urlprefix  [0]{URL }%
\providecommand \Eprint [0]{\href }%
\providecommand \doibase [0]{http://dx.doi.org/}%
\providecommand \selectlanguage [0]{\@gobble}%
\providecommand \bibinfo  [0]{\@secondoftwo}%
\providecommand \bibfield  [0]{\@secondoftwo}%
\providecommand \translation [1]{[#1]}%
\providecommand \BibitemOpen [0]{}%
\providecommand \bibitemStop [0]{}%
\providecommand \bibitemNoStop [0]{.\EOS\space}%
\providecommand \EOS [0]{\spacefactor3000\relax}%
\providecommand \BibitemShut  [1]{\csname bibitem#1\endcsname}%
\let\auto@bib@innerbib\@empty
\bibitem [{\citenamefont {Flaxman}\ \emph {et~al.}(2020)\citenamefont
  {Flaxman}, \citenamefont {Mishra}, \citenamefont {Gandy}, \citenamefont
  {Unwin}, \citenamefont {Mellan}, \citenamefont {Coupland}, \citenamefont
  {Whittaker}, \citenamefont {Zhu}, \citenamefont {Berah}, \citenamefont
  {Eaton} \emph {et~al.}}]{flaxman2020estimating}%
  \BibitemOpen
  \bibfield  {author} {\bibinfo {author} {\bibfnamefont {S.}~\bibnamefont
  {Flaxman}}, \bibinfo {author} {\bibfnamefont {S.}~\bibnamefont {Mishra}},
  \bibinfo {author} {\bibfnamefont {A.}~\bibnamefont {Gandy}}, \bibinfo
  {author} {\bibfnamefont {H.~J.~T.}\ \bibnamefont {Unwin}}, \bibinfo {author}
  {\bibfnamefont {T.~A.}\ \bibnamefont {Mellan}}, \bibinfo {author}
  {\bibfnamefont {H.}~\bibnamefont {Coupland}}, \bibinfo {author}
  {\bibfnamefont {C.}~\bibnamefont {Whittaker}}, \bibinfo {author}
  {\bibfnamefont {H.}~\bibnamefont {Zhu}}, \bibinfo {author} {\bibfnamefont
  {T.}~\bibnamefont {Berah}}, \bibinfo {author} {\bibfnamefont {J.~W.}\
  \bibnamefont {Eaton}},  \emph {et~al.},\ }\href@noop {} {\bibfield  {journal}
  {\bibinfo  {journal} {Nature}\ }\textbf {\bibinfo {volume} {584}},\ \bibinfo
  {pages} {257} (\bibinfo {year} {2020})}\BibitemShut {NoStop}%
\bibitem [{\citenamefont {Fong}\ \emph {et~al.}(2020)\citenamefont {Fong},
  \citenamefont {Gao}, \citenamefont {Wong}, \citenamefont {Xiao},
  \citenamefont {Shiu}, \citenamefont {Ryu},\ and\ \citenamefont
  {Cowling}}]{fong2020nonpharmaceutical}%
  \BibitemOpen
  \bibfield  {author} {\bibinfo {author} {\bibfnamefont {M.~W.}\ \bibnamefont
  {Fong}}, \bibinfo {author} {\bibfnamefont {H.}~\bibnamefont {Gao}}, \bibinfo
  {author} {\bibfnamefont {J.~Y.}\ \bibnamefont {Wong}}, \bibinfo {author}
  {\bibfnamefont {J.}~\bibnamefont {Xiao}}, \bibinfo {author} {\bibfnamefont
  {E.~Y.}\ \bibnamefont {Shiu}}, \bibinfo {author} {\bibfnamefont
  {S.}~\bibnamefont {Ryu}}, \ and\ \bibinfo {author} {\bibfnamefont {B.~J.}\
  \bibnamefont {Cowling}},\ }\href@noop {} {\bibfield  {journal} {\bibinfo
  {journal} {Emerg. Infect. Dis.}\ }\textbf {\bibinfo {volume} {26}},\ \bibinfo
  {pages} {976} (\bibinfo {year} {2020})}\BibitemShut {NoStop}%
\bibitem [{\citenamefont {Markel}\ \emph {et~al.}(2007)\citenamefont {Markel},
  \citenamefont {Lipman}, \citenamefont {Navarro}, \citenamefont {Sloan},
  \citenamefont {Michalsen}, \citenamefont {Stern},\ and\ \citenamefont
  {Cetron}}]{markel2007nonpharmaceutical}%
  \BibitemOpen
  \bibfield  {author} {\bibinfo {author} {\bibfnamefont {H.}~\bibnamefont
  {Markel}}, \bibinfo {author} {\bibfnamefont {H.~B.}\ \bibnamefont {Lipman}},
  \bibinfo {author} {\bibfnamefont {J.~A.}\ \bibnamefont {Navarro}}, \bibinfo
  {author} {\bibfnamefont {A.}~\bibnamefont {Sloan}}, \bibinfo {author}
  {\bibfnamefont {J.~R.}\ \bibnamefont {Michalsen}}, \bibinfo {author}
  {\bibfnamefont {A.~M.}\ \bibnamefont {Stern}}, \ and\ \bibinfo {author}
  {\bibfnamefont {M.~S.}\ \bibnamefont {Cetron}},\ }\href@noop {} {\bibfield
  {journal} {\bibinfo  {journal} {JAMA}\ }\textbf {\bibinfo {volume} {298}},\
  \bibinfo {pages} {644} (\bibinfo {year} {2007})}\BibitemShut {NoStop}%
\bibitem [{\citenamefont {De}\ \emph {et~al.}(2004)\citenamefont {De},
  \citenamefont {Singh}, \citenamefont {Wong}, \citenamefont {Yacoub},\ and\
  \citenamefont {Jolly}}]{de2004sexual}%
  \BibitemOpen
  \bibfield  {author} {\bibinfo {author} {\bibfnamefont {P.}~\bibnamefont
  {De}}, \bibinfo {author} {\bibfnamefont {A.~E.}\ \bibnamefont {Singh}},
  \bibinfo {author} {\bibfnamefont {T.}~\bibnamefont {Wong}}, \bibinfo {author}
  {\bibfnamefont {W.}~\bibnamefont {Yacoub}}, \ and\ \bibinfo {author}
  {\bibfnamefont {A.}~\bibnamefont {Jolly}},\ }\href@noop {} {\bibfield
  {journal} {\bibinfo  {journal} {Sex. Transm. Infect.}\ }\textbf {\bibinfo
  {volume} {80}},\ \bibinfo {pages} {280} (\bibinfo {year} {2004})}\BibitemShut
  {NoStop}%
\bibitem [{\citenamefont {Susswein}\ and\ \citenamefont
  {Bansal}(2020)}]{susswein2020characterizing}%
  \BibitemOpen
  \bibfield  {author} {\bibinfo {author} {\bibfnamefont {Z.}~\bibnamefont
  {Susswein}}\ and\ \bibinfo {author} {\bibfnamefont {S.}~\bibnamefont
  {Bansal}},\ }\href@noop {} {\bibfield  {journal} {\bibinfo  {journal}
  {MedRxiv}\ } (\bibinfo {year} {2020})}\BibitemShut {NoStop}%
\bibitem [{\citenamefont {Luke}\ and\ \citenamefont
  {Harris}(2007)}]{luke2007network}%
  \BibitemOpen
  \bibfield  {author} {\bibinfo {author} {\bibfnamefont {D.~A.}\ \bibnamefont
  {Luke}}\ and\ \bibinfo {author} {\bibfnamefont {J.~K.}\ \bibnamefont
  {Harris}},\ }\href@noop {} {\bibfield  {journal} {\bibinfo  {journal} {Annu.
  Rev. Public Health}\ }\textbf {\bibinfo {volume} {28}},\ \bibinfo {pages}
  {69} (\bibinfo {year} {2007})}\BibitemShut {NoStop}%
\bibitem [{\citenamefont {Wang}\ \emph {et~al.}(2015)\citenamefont {Wang},
  \citenamefont {Andrews}, \citenamefont {Wu}, \citenamefont {Wang},\ and\
  \citenamefont {Bauch}}]{wang2015coupled}%
  \BibitemOpen
  \bibfield  {author} {\bibinfo {author} {\bibfnamefont {Z.}~\bibnamefont
  {Wang}}, \bibinfo {author} {\bibfnamefont {M.~A.}\ \bibnamefont {Andrews}},
  \bibinfo {author} {\bibfnamefont {Z.-X.}\ \bibnamefont {Wu}}, \bibinfo
  {author} {\bibfnamefont {L.}~\bibnamefont {Wang}}, \ and\ \bibinfo {author}
  {\bibfnamefont {C.~T.}\ \bibnamefont {Bauch}},\ }\href@noop {} {\bibfield
  {journal} {\bibinfo  {journal} {Phys. Life Rev.}\ }\textbf {\bibinfo {volume}
  {15}},\ \bibinfo {pages} {1} (\bibinfo {year} {2015})}\BibitemShut {NoStop}%
\bibitem [{\citenamefont {Wang}\ \emph {et~al.}(2021)\citenamefont {Wang},
  \citenamefont {An}, \citenamefont {He},\ and\ \citenamefont
  {Fang}}]{wang2021literature}%
  \BibitemOpen
  \bibfield  {author} {\bibinfo {author} {\bibfnamefont {X.}~\bibnamefont
  {Wang}}, \bibinfo {author} {\bibfnamefont {Q.}~\bibnamefont {An}}, \bibinfo
  {author} {\bibfnamefont {Z.}~\bibnamefont {He}}, \ and\ \bibinfo {author}
  {\bibfnamefont {W.}~\bibnamefont {Fang}},\ }\href@noop {} {\bibfield
  {journal} {\bibinfo  {journal} {Complexity}\ }\textbf {\bibinfo {volume}
  {2021}} (\bibinfo {year} {2021})}\BibitemShut {NoStop}%
\bibitem [{\citenamefont {Kojaku}\ \emph {et~al.}(2021)\citenamefont {Kojaku},
  \citenamefont {H{\'e}bert-Dufresne}, \citenamefont {Mones}, \citenamefont
  {Lehmann},\ and\ \citenamefont {Ahn}}]{kojaku2021effectiveness}%
  \BibitemOpen
  \bibfield  {author} {\bibinfo {author} {\bibfnamefont {S.}~\bibnamefont
  {Kojaku}}, \bibinfo {author} {\bibfnamefont {L.}~\bibnamefont
  {H{\'e}bert-Dufresne}}, \bibinfo {author} {\bibfnamefont {E.}~\bibnamefont
  {Mones}}, \bibinfo {author} {\bibfnamefont {S.}~\bibnamefont {Lehmann}}, \
  and\ \bibinfo {author} {\bibfnamefont {Y.-Y.}\ \bibnamefont {Ahn}},\
  }\href@noop {} {\bibfield  {journal} {\bibinfo  {journal} {Nat. Phys.}\
  }\textbf {\bibinfo {volume} {17}},\ \bibinfo {pages} {652} (\bibinfo {year}
  {2021})}\BibitemShut {NoStop}%
\bibitem [{\citenamefont {Rizi}\ \emph {et~al.}(2022)\citenamefont {Rizi},
  \citenamefont {Faqeeh}, \citenamefont {Badie-Modiri},\ and\ \citenamefont
  {Kivel{\"a}}}]{rizi2022epidemic}%
  \BibitemOpen
  \bibfield  {author} {\bibinfo {author} {\bibfnamefont {A.~K.}\ \bibnamefont
  {Rizi}}, \bibinfo {author} {\bibfnamefont {A.}~\bibnamefont {Faqeeh}},
  \bibinfo {author} {\bibfnamefont {A.}~\bibnamefont {Badie-Modiri}}, \ and\
  \bibinfo {author} {\bibfnamefont {M.}~\bibnamefont {Kivel{\"a}}},\
  }\href@noop {} {\bibfield  {journal} {\bibinfo  {journal} {Phys. Rev. E}\
  }\textbf {\bibinfo {volume} {105}},\ \bibinfo {pages} {044313} (\bibinfo
  {year} {2022})}\BibitemShut {NoStop}%
\bibitem [{\citenamefont {St-Onge}\ \emph
  {et~al.}(2021{\natexlab{a}})\citenamefont {St-Onge}, \citenamefont
  {Thibeault}, \citenamefont {Allard}, \citenamefont {Dub{\'e}},\ and\
  \citenamefont {H{\'e}bert-Dufresne}}]{st2021social}%
  \BibitemOpen
  \bibfield  {author} {\bibinfo {author} {\bibfnamefont {G.}~\bibnamefont
  {St-Onge}}, \bibinfo {author} {\bibfnamefont {V.}~\bibnamefont {Thibeault}},
  \bibinfo {author} {\bibfnamefont {A.}~\bibnamefont {Allard}}, \bibinfo
  {author} {\bibfnamefont {L.~J.}\ \bibnamefont {Dub{\'e}}}, \ and\ \bibinfo
  {author} {\bibfnamefont {L.}~\bibnamefont {H{\'e}bert-Dufresne}},\
  }\href@noop {} {\bibfield  {journal} {\bibinfo  {journal} {Phys. Rev. Lett.}\
  }\textbf {\bibinfo {volume} {126}},\ \bibinfo {pages} {098301} (\bibinfo
  {year} {2021}{\natexlab{a}})}\BibitemShut {NoStop}%
\bibitem [{\citenamefont {St-Onge}\ \emph
  {et~al.}(2021{\natexlab{b}})\citenamefont {St-Onge}, \citenamefont
  {Thibeault}, \citenamefont {Allard}, \citenamefont {Dub{\'e}},\ and\
  \citenamefont {H{\'e}bert-Dufresne}}]{st2021master}%
  \BibitemOpen
  \bibfield  {author} {\bibinfo {author} {\bibfnamefont {G.}~\bibnamefont
  {St-Onge}}, \bibinfo {author} {\bibfnamefont {V.}~\bibnamefont {Thibeault}},
  \bibinfo {author} {\bibfnamefont {A.}~\bibnamefont {Allard}}, \bibinfo
  {author} {\bibfnamefont {L.~J.}\ \bibnamefont {Dub{\'e}}}, \ and\ \bibinfo
  {author} {\bibfnamefont {L.}~\bibnamefont {H{\'e}bert-Dufresne}},\
  }\href@noop {} {\bibfield  {journal} {\bibinfo  {journal} {Phys. Rev. E}\
  }\textbf {\bibinfo {volume} {103}},\ \bibinfo {pages} {032301} (\bibinfo
  {year} {2021}{\natexlab{b}})}\BibitemShut {NoStop}%
\bibitem [{\citenamefont {Gross}\ \emph {et~al.}(2006)\citenamefont {Gross},
  \citenamefont {D’Lima},\ and\ \citenamefont {Blasius}}]{gross2006epidemic}%
  \BibitemOpen
  \bibfield  {author} {\bibinfo {author} {\bibfnamefont {T.}~\bibnamefont
  {Gross}}, \bibinfo {author} {\bibfnamefont {C.~J.~D.}\ \bibnamefont
  {D’Lima}}, \ and\ \bibinfo {author} {\bibfnamefont {B.}~\bibnamefont
  {Blasius}},\ }\href@noop {} {\bibfield  {journal} {\bibinfo  {journal} {Phys.
  Rev. Lett.}\ }\textbf {\bibinfo {volume} {96}},\ \bibinfo {pages} {208701}
  (\bibinfo {year} {2006})}\BibitemShut {NoStop}%
\bibitem [{\citenamefont {Durrett}\ and\ \citenamefont
  {Yao}(2022)}]{durrett2022susceptible}%
  \BibitemOpen
  \bibfield  {author} {\bibinfo {author} {\bibfnamefont {R.}~\bibnamefont
  {Durrett}}\ and\ \bibinfo {author} {\bibfnamefont {D.}~\bibnamefont {Yao}},\
  }\href@noop {} {\bibfield  {journal} {\bibinfo  {journal} {Electron. J.
  Probab.}\ }\textbf {\bibinfo {volume} {27}},\ \bibinfo {pages} {1} (\bibinfo
  {year} {2022})}\BibitemShut {NoStop}%
\bibitem [{\citenamefont {Ball}\ and\ \citenamefont
  {Britton}(2022)}]{ball2022epidemics}%
  \BibitemOpen
  \bibfield  {author} {\bibinfo {author} {\bibfnamefont {F.}~\bibnamefont
  {Ball}}\ and\ \bibinfo {author} {\bibfnamefont {T.}~\bibnamefont {Britton}},\
  }\href@noop {} {\bibfield  {journal} {\bibinfo  {journal} {Random Structures
  \& Algorithms}\ }\textbf {\bibinfo {volume} {61}},\ \bibinfo {pages} {250}
  (\bibinfo {year} {2022})}\BibitemShut {NoStop}%
\bibitem [{\citenamefont {Strona}\ and\ \citenamefont
  {Castellano}(2018)}]{strona2018rapid}%
  \BibitemOpen
  \bibfield  {author} {\bibinfo {author} {\bibfnamefont {G.}~\bibnamefont
  {Strona}}\ and\ \bibinfo {author} {\bibfnamefont {C.}~\bibnamefont
  {Castellano}},\ }\href@noop {} {\bibfield  {journal} {\bibinfo  {journal}
  {Phys. Rev. E}\ }\textbf {\bibinfo {volume} {97}},\ \bibinfo {pages} {022308}
  (\bibinfo {year} {2018})}\BibitemShut {NoStop}%
\bibitem [{\citenamefont {Vyasarayani}\ and\ \citenamefont
  {Chatterjee}(2020)}]{vyasarayani2020complete}%
  \BibitemOpen
  \bibfield  {author} {\bibinfo {author} {\bibfnamefont {C.}~\bibnamefont
  {Vyasarayani}}\ and\ \bibinfo {author} {\bibfnamefont {A.}~\bibnamefont
  {Chatterjee}},\ }\href@noop {} {\bibfield  {journal} {\bibinfo  {journal}
  {Nonlinear Dyn.}\ }\textbf {\bibinfo {volume} {101}},\ \bibinfo {pages}
  {1653} (\bibinfo {year} {2020})}\BibitemShut {NoStop}%
\bibitem [{\citenamefont {Hasegawa}\ and\ \citenamefont
  {Nemoto}(2017)}]{hasegawa2017efficiency}%
  \BibitemOpen
  \bibfield  {author} {\bibinfo {author} {\bibfnamefont {T.}~\bibnamefont
  {Hasegawa}}\ and\ \bibinfo {author} {\bibfnamefont {K.}~\bibnamefont
  {Nemoto}},\ }\href@noop {} {\bibfield  {journal} {\bibinfo  {journal} {Phys.
  Rev. e}\ }\textbf {\bibinfo {volume} {96}},\ \bibinfo {pages} {022311}
  (\bibinfo {year} {2017})}\BibitemShut {NoStop}%
\bibitem [{\citenamefont {B{\"o}rner}\ \emph {et~al.}(2022)\citenamefont
  {B{\"o}rner}, \citenamefont {Schr{\"o}der}, \citenamefont {Scarselli},
  \citenamefont {Budanur}, \citenamefont {Hof},\ and\ \citenamefont
  {Timme}}]{borner2022explosive}%
  \BibitemOpen
  \bibfield  {author} {\bibinfo {author} {\bibfnamefont {G.}~\bibnamefont
  {B{\"o}rner}}, \bibinfo {author} {\bibfnamefont {M.}~\bibnamefont
  {Schr{\"o}der}}, \bibinfo {author} {\bibfnamefont {D.}~\bibnamefont
  {Scarselli}}, \bibinfo {author} {\bibfnamefont {N.~B.}\ \bibnamefont
  {Budanur}}, \bibinfo {author} {\bibfnamefont {B.}~\bibnamefont {Hof}}, \ and\
  \bibinfo {author} {\bibfnamefont {M.}~\bibnamefont {Timme}},\ }\href@noop {}
  {\bibfield  {journal} {\bibinfo  {journal} {J. Phys. Complex.}\ }\textbf
  {\bibinfo {volume} {3}},\ \bibinfo {pages} {04LT02} (\bibinfo {year}
  {2022})}\BibitemShut {NoStop}%
\bibitem [{\citenamefont {Molloy}\ and\ \citenamefont
  {Reed}(1998)}]{molloy1998size}%
  \BibitemOpen
  \bibfield  {author} {\bibinfo {author} {\bibfnamefont {M.}~\bibnamefont
  {Molloy}}\ and\ \bibinfo {author} {\bibfnamefont {B.}~\bibnamefont {Reed}},\
  }\href@noop {} {\bibfield  {journal} {\bibinfo  {journal} {Comb. Probab.
  Comput.}\ }\textbf {\bibinfo {volume} {7}},\ \bibinfo {pages} {295} (\bibinfo
  {year} {1998})}\BibitemShut {NoStop}%
\bibitem [{\citenamefont {Molloy}\ and\ \citenamefont
  {Reed}(1995)}]{molloy1995critical}%
  \BibitemOpen
  \bibfield  {author} {\bibinfo {author} {\bibfnamefont {M.}~\bibnamefont
  {Molloy}}\ and\ \bibinfo {author} {\bibfnamefont {B.}~\bibnamefont {Reed}},\
  }\href@noop {} {\bibfield  {journal} {\bibinfo  {journal} {Random structures
  \& algorithms}\ }\textbf {\bibinfo {volume} {6}},\ \bibinfo {pages} {161}
  (\bibinfo {year} {1995})}\BibitemShut {NoStop}%
\bibitem [{\citenamefont {Karrer}\ and\ \citenamefont
  {Newman}(2010)}]{karrer2010random}%
  \BibitemOpen
  \bibfield  {author} {\bibinfo {author} {\bibfnamefont {B.}~\bibnamefont
  {Karrer}}\ and\ \bibinfo {author} {\bibfnamefont {M.~E.}\ \bibnamefont
  {Newman}},\ }\href@noop {} {\bibfield  {journal} {\bibinfo  {journal} {Phys.
  Rev. E}\ }\textbf {\bibinfo {volume} {82}},\ \bibinfo {pages} {066118}
  (\bibinfo {year} {2010})}\BibitemShut {NoStop}%
\bibitem [{\citenamefont {Newman}(2002)}]{newman2002spread}%
  \BibitemOpen
  \bibfield  {author} {\bibinfo {author} {\bibfnamefont {M.~E.}\ \bibnamefont
  {Newman}},\ }\href@noop {} {\bibfield  {journal} {\bibinfo  {journal} {Phys.
  Rev. E}\ }\textbf {\bibinfo {volume} {66}},\ \bibinfo {pages} {016128}
  (\bibinfo {year} {2002})}\BibitemShut {NoStop}%
\bibitem [{\citenamefont {Radicchi}\ and\ \citenamefont
  {Bianconi}(2020)}]{radicchi2020epidemic}%
  \BibitemOpen
  \bibfield  {author} {\bibinfo {author} {\bibfnamefont {F.}~\bibnamefont
  {Radicchi}}\ and\ \bibinfo {author} {\bibfnamefont {G.}~\bibnamefont
  {Bianconi}},\ }\href@noop {} {\bibfield  {journal} {\bibinfo  {journal}
  {Phys. Rev. E}\ }\textbf {\bibinfo {volume} {102}},\ \bibinfo {pages}
  {052309} (\bibinfo {year} {2020})}\BibitemShut {NoStop}%
\bibitem [{\citenamefont {Miller}(2014)}]{miller2014epidemics}%
  \BibitemOpen
  \bibfield  {author} {\bibinfo {author} {\bibfnamefont {J.~C.}\ \bibnamefont
  {Miller}},\ }\href@noop {} {\bibfield  {journal} {\bibinfo  {journal} {PloS
  One}\ }\textbf {\bibinfo {volume} {9}},\ \bibinfo {pages} {e101421} (\bibinfo
  {year} {2014})}\BibitemShut {NoStop}%
\bibitem [{\citenamefont {Krapivsky}(2021)}]{krapivsky2021infection}%
  \BibitemOpen
  \bibfield  {author} {\bibinfo {author} {\bibfnamefont {P.}~\bibnamefont
  {Krapivsky}},\ }\href@noop {} {\bibfield  {journal} {\bibinfo  {journal} {J.
  Stat. Mech.: Theory Exp.}\ }\textbf {\bibinfo {volume} {2021}},\ \bibinfo
  {pages} {013501} (\bibinfo {year} {2021})}\BibitemShut {NoStop}%
\bibitem [{\citenamefont {Machado}\ and\ \citenamefont
  {Baxter}(2022)}]{machado2022effect}%
  \BibitemOpen
  \bibfield  {author} {\bibinfo {author} {\bibfnamefont {G.}~\bibnamefont
  {Machado}}\ and\ \bibinfo {author} {\bibfnamefont {G.}~\bibnamefont
  {Baxter}},\ }\href@noop {} {\bibfield  {journal} {\bibinfo  {journal} {Phys.
  Rev. E}\ }\textbf {\bibinfo {volume} {106}},\ \bibinfo {pages} {014307}
  (\bibinfo {year} {2022})}\BibitemShut {NoStop}%
\bibitem [{\citenamefont {Hasegawa}\ and\ \citenamefont
  {Nemoto}(2016)}]{hasegawa2016outbreaks}%
  \BibitemOpen
  \bibfield  {author} {\bibinfo {author} {\bibfnamefont {T.}~\bibnamefont
  {Hasegawa}}\ and\ \bibinfo {author} {\bibfnamefont {K.}~\bibnamefont
  {Nemoto}},\ }\href@noop {} {\bibfield  {journal} {\bibinfo  {journal} {Phys.
  Rev. e}\ }\textbf {\bibinfo {volume} {93}},\ \bibinfo {pages} {032324}
  (\bibinfo {year} {2016})}\BibitemShut {NoStop}%
\bibitem [{\citenamefont {Brauer}\ \emph {et~al.}(2008)\citenamefont {Brauer},
  \citenamefont {Van~den Driessche}, \citenamefont {Wu},\ and\ \citenamefont
  {Allen}}]{brauer2008mathematical}%
  \BibitemOpen
  \bibfield  {author} {\bibinfo {author} {\bibfnamefont {F.}~\bibnamefont
  {Brauer}}, \bibinfo {author} {\bibfnamefont {P.}~\bibnamefont {Van~den
  Driessche}}, \bibinfo {author} {\bibfnamefont {J.}~\bibnamefont {Wu}}, \ and\
  \bibinfo {author} {\bibfnamefont {L.~J.}\ \bibnamefont {Allen}},\ }\href@noop
  {} {\emph {\bibinfo {title} {Mathematical epidemiology}}},\ Vol.\ \bibinfo
  {volume} {1945}\ (\bibinfo  {publisher} {Springer},\ \bibinfo {year}
  {2008})\BibitemShut {NoStop}%
\bibitem [{\citenamefont {Stauffer}\ and\ \citenamefont
  {Aharony}(2014)}]{stauffer2014introduction}%
  \BibitemOpen
  \bibfield  {author} {\bibinfo {author} {\bibfnamefont {D.}~\bibnamefont
  {Stauffer}}\ and\ \bibinfo {author} {\bibfnamefont {A.}~\bibnamefont
  {Aharony}},\ }\href@noop {} {\emph {\bibinfo {title} {Introduction to
  percolation theory: revised second edition}}}\ (\bibinfo  {publisher} {CRC
  press},\ \bibinfo {year} {2014})\BibitemShut {NoStop}%
\bibitem [{\citenamefont {Grassberger}(1983)}]{grassberger1983critical}%
  \BibitemOpen
  \bibfield  {author} {\bibinfo {author} {\bibfnamefont {P.}~\bibnamefont
  {Grassberger}},\ }\href@noop {} {\bibfield  {journal} {\bibinfo  {journal}
  {Math. Biosci.}\ }\textbf {\bibinfo {volume} {63}},\ \bibinfo {pages} {157}
  (\bibinfo {year} {1983})}\BibitemShut {NoStop}%
\bibitem [{Exx()}]{Exxon01}%
  \BibitemOpen
  \href@noop {} {}\bibinfo {note} {It is worth noting that the process under
  study is non-Markovian and infected individuals recover with probability 1
  after $t_r$ time steps.}\BibitemShut {Stop}%
\bibitem [{\citenamefont {Battiston}\ \emph {et~al.}(2020)\citenamefont
  {Battiston}, \citenamefont {Cencetti}, \citenamefont {Iacopini},
  \citenamefont {Latora}, \citenamefont {Lucas}, \citenamefont {Patania},
  \citenamefont {Young},\ and\ \citenamefont {Petri}}]{battiston2020networks}%
  \BibitemOpen
  \bibfield  {author} {\bibinfo {author} {\bibfnamefont {F.}~\bibnamefont
  {Battiston}}, \bibinfo {author} {\bibfnamefont {G.}~\bibnamefont {Cencetti}},
  \bibinfo {author} {\bibfnamefont {I.}~\bibnamefont {Iacopini}}, \bibinfo
  {author} {\bibfnamefont {V.}~\bibnamefont {Latora}}, \bibinfo {author}
  {\bibfnamefont {M.}~\bibnamefont {Lucas}}, \bibinfo {author} {\bibfnamefont
  {A.}~\bibnamefont {Patania}}, \bibinfo {author} {\bibfnamefont {J.-G.}\
  \bibnamefont {Young}}, \ and\ \bibinfo {author} {\bibfnamefont
  {G.}~\bibnamefont {Petri}},\ }\href@noop {} {\bibfield  {journal} {\bibinfo
  {journal} {Phys. Rep.}\ }\textbf {\bibinfo {volume} {874}},\ \bibinfo {pages}
  {1} (\bibinfo {year} {2020})}\BibitemShut {NoStop}%
\bibitem [{\citenamefont {Battiston}\ \emph {et~al.}(2021)\citenamefont
  {Battiston}, \citenamefont {Amico}, \citenamefont {Barrat}, \citenamefont
  {Bianconi}, \citenamefont {Ferraz~de Arruda}, \citenamefont {Franceschiello},
  \citenamefont {Iacopini}, \citenamefont {K{\'e}fi}, \citenamefont {Latora},
  \citenamefont {Moreno} \emph {et~al.}}]{battiston2021physics}%
  \BibitemOpen
  \bibfield  {author} {\bibinfo {author} {\bibfnamefont {F.}~\bibnamefont
  {Battiston}}, \bibinfo {author} {\bibfnamefont {E.}~\bibnamefont {Amico}},
  \bibinfo {author} {\bibfnamefont {A.}~\bibnamefont {Barrat}}, \bibinfo
  {author} {\bibfnamefont {G.}~\bibnamefont {Bianconi}}, \bibinfo {author}
  {\bibfnamefont {G.}~\bibnamefont {Ferraz~de Arruda}}, \bibinfo {author}
  {\bibfnamefont {B.}~\bibnamefont {Franceschiello}}, \bibinfo {author}
  {\bibfnamefont {I.}~\bibnamefont {Iacopini}}, \bibinfo {author}
  {\bibfnamefont {S.}~\bibnamefont {K{\'e}fi}}, \bibinfo {author}
  {\bibfnamefont {V.}~\bibnamefont {Latora}}, \bibinfo {author} {\bibfnamefont
  {Y.}~\bibnamefont {Moreno}},  \emph {et~al.},\ }\href@noop {} {\bibfield
  {journal} {\bibinfo  {journal} {Nat. Phys.}\ }\textbf {\bibinfo {volume}
  {17}},\ \bibinfo {pages} {1093} (\bibinfo {year} {2021})}\BibitemShut
  {NoStop}%
\bibitem [{\citenamefont {Kenah}\ and\ \citenamefont
  {Robins}(2007)}]{kenah2007second}%
  \BibitemOpen
  \bibfield  {author} {\bibinfo {author} {\bibfnamefont {E.}~\bibnamefont
  {Kenah}}\ and\ \bibinfo {author} {\bibfnamefont {J.~M.}\ \bibnamefont
  {Robins}},\ }\href@noop {} {\bibfield  {journal} {\bibinfo  {journal} {Phys.
  Rev. E}\ }\textbf {\bibinfo {volume} {76}},\ \bibinfo {pages} {036113}
  (\bibinfo {year} {2007})}\BibitemShut {NoStop}%
\bibitem [{\citenamefont {Meyers}\ \emph {et~al.}(2006)\citenamefont {Meyers},
  \citenamefont {Newman},\ and\ \citenamefont
  {Pourbohloul}}]{meyers2006predicting}%
  \BibitemOpen
  \bibfield  {author} {\bibinfo {author} {\bibfnamefont {L.~A.}\ \bibnamefont
  {Meyers}}, \bibinfo {author} {\bibfnamefont {M.}~\bibnamefont {Newman}}, \
  and\ \bibinfo {author} {\bibfnamefont {B.}~\bibnamefont {Pourbohloul}},\
  }\href@noop {} {\bibfield  {journal} {\bibinfo  {journal} {J. Theor. Biol.}\
  }\textbf {\bibinfo {volume} {240}},\ \bibinfo {pages} {400} (\bibinfo {year}
  {2006})}\BibitemShut {NoStop}%
\bibitem [{\citenamefont {Miller}(2009)}]{miller2009spread}%
  \BibitemOpen
  \bibfield  {author} {\bibinfo {author} {\bibfnamefont {J.~C.}\ \bibnamefont
  {Miller}},\ }\href@noop {} {\bibfield  {journal} {\bibinfo  {journal} {J. R.
  Soc. Interface}\ }\textbf {\bibinfo {volume} {6}},\ \bibinfo {pages} {1121}
  (\bibinfo {year} {2009})}\BibitemShut {NoStop}%
\bibitem [{\citenamefont {Valdez}\ \emph
  {et~al.}(2020{\natexlab{a}})\citenamefont {Valdez}, \citenamefont
  {Braunstein},\ and\ \citenamefont {Havlin}}]{valdez2020epidemic}%
  \BibitemOpen
  \bibfield  {author} {\bibinfo {author} {\bibfnamefont {L.~D.}\ \bibnamefont
  {Valdez}}, \bibinfo {author} {\bibfnamefont {L.~A.}\ \bibnamefont
  {Braunstein}}, \ and\ \bibinfo {author} {\bibfnamefont {S.}~\bibnamefont
  {Havlin}},\ }\href@noop {} {\bibfield  {journal} {\bibinfo  {journal} {Phys.
  Rev. E}\ }\textbf {\bibinfo {volume} {101}},\ \bibinfo {pages} {032309}
  (\bibinfo {year} {2020}{\natexlab{a}})}\BibitemShut {NoStop}%
\bibitem [{\citenamefont {Buldyrev}\ \emph {et~al.}(2010)\citenamefont
  {Buldyrev}, \citenamefont {Parshani}, \citenamefont {Paul}, \citenamefont
  {Stanley},\ and\ \citenamefont {Havlin}}]{buldyrev2010catastrophic}%
  \BibitemOpen
  \bibfield  {author} {\bibinfo {author} {\bibfnamefont {S.~V.}\ \bibnamefont
  {Buldyrev}}, \bibinfo {author} {\bibfnamefont {R.}~\bibnamefont {Parshani}},
  \bibinfo {author} {\bibfnamefont {G.}~\bibnamefont {Paul}}, \bibinfo {author}
  {\bibfnamefont {H.~E.}\ \bibnamefont {Stanley}}, \ and\ \bibinfo {author}
  {\bibfnamefont {S.}~\bibnamefont {Havlin}},\ }\href@noop {} {\bibfield
  {journal} {\bibinfo  {journal} {Nature}\ }\textbf {\bibinfo {volume} {464}},\
  \bibinfo {pages} {1025} (\bibinfo {year} {2010})}\BibitemShut {NoStop}%
\bibitem [{\citenamefont {Valdez}\ \emph
  {et~al.}(2020{\natexlab{b}})\citenamefont {Valdez}, \citenamefont
  {Shekhtman}, \citenamefont {La~Rocca}, \citenamefont {Zhang}, \citenamefont
  {Buldyrev}, \citenamefont {Trunfio}, \citenamefont {Braunstein},\ and\
  \citenamefont {Havlin}}]{valdez2020cascading}%
  \BibitemOpen
  \bibfield  {author} {\bibinfo {author} {\bibfnamefont {L.~D.}\ \bibnamefont
  {Valdez}}, \bibinfo {author} {\bibfnamefont {L.}~\bibnamefont {Shekhtman}},
  \bibinfo {author} {\bibfnamefont {C.~E.}\ \bibnamefont {La~Rocca}}, \bibinfo
  {author} {\bibfnamefont {X.}~\bibnamefont {Zhang}}, \bibinfo {author}
  {\bibfnamefont {S.~V.}\ \bibnamefont {Buldyrev}}, \bibinfo {author}
  {\bibfnamefont {P.~A.}\ \bibnamefont {Trunfio}}, \bibinfo {author}
  {\bibfnamefont {L.~A.}\ \bibnamefont {Braunstein}}, \ and\ \bibinfo {author}
  {\bibfnamefont {S.}~\bibnamefont {Havlin}},\ }\href@noop {} {\bibfield
  {journal} {\bibinfo  {journal} {J. Complex Netw.}\ }\textbf {\bibinfo
  {volume} {8}},\ \bibinfo {pages} {cnaa013} (\bibinfo {year}
  {2020}{\natexlab{b}})}\BibitemShut {NoStop}%
\bibitem [{\citenamefont {Pastor-Satorras}\ \emph {et~al.}(2015)\citenamefont
  {Pastor-Satorras}, \citenamefont {Castellano}, \citenamefont {Van~Mieghem},\
  and\ \citenamefont {Vespignani}}]{pastor2015epidemic}%
  \BibitemOpen
  \bibfield  {author} {\bibinfo {author} {\bibfnamefont {R.}~\bibnamefont
  {Pastor-Satorras}}, \bibinfo {author} {\bibfnamefont {C.}~\bibnamefont
  {Castellano}}, \bibinfo {author} {\bibfnamefont {P.}~\bibnamefont
  {Van~Mieghem}}, \ and\ \bibinfo {author} {\bibfnamefont {A.}~\bibnamefont
  {Vespignani}},\ }\href@noop {} {\bibfield  {journal} {\bibinfo  {journal}
  {Rev. Mod. Phys.}\ }\textbf {\bibinfo {volume} {87}},\ \bibinfo {pages} {925}
  (\bibinfo {year} {2015})}\BibitemShut {NoStop}%
\bibitem [{\citenamefont {Wang}\ \emph {et~al.}(2018)\citenamefont {Wang},
  \citenamefont {Wang},\ and\ \citenamefont {Cai}}]{wang2018critical}%
  \BibitemOpen
  \bibfield  {author} {\bibinfo {author} {\bibfnamefont {W.}~\bibnamefont
  {Wang}}, \bibinfo {author} {\bibfnamefont {Z.-X.}\ \bibnamefont {Wang}}, \
  and\ \bibinfo {author} {\bibfnamefont {S.-M.}\ \bibnamefont {Cai}},\
  }\href@noop {} {\bibfield  {journal} {\bibinfo  {journal} {Phys. Rev. E}\
  }\textbf {\bibinfo {volume} {98}},\ \bibinfo {pages} {052312} (\bibinfo
  {year} {2018})}\BibitemShut {NoStop}%
\bibitem [{\citenamefont {Dong}\ \emph {et~al.}(2021)\citenamefont {Dong},
  \citenamefont {Wang}, \citenamefont {Shekhtman}, \citenamefont {Danziger},
  \citenamefont {Fan}, \citenamefont {Du}, \citenamefont {Liu}, \citenamefont
  {Tian}, \citenamefont {Stanley},\ and\ \citenamefont
  {Havlin}}]{dong2021optimal}%
  \BibitemOpen
  \bibfield  {author} {\bibinfo {author} {\bibfnamefont {G.}~\bibnamefont
  {Dong}}, \bibinfo {author} {\bibfnamefont {F.}~\bibnamefont {Wang}}, \bibinfo
  {author} {\bibfnamefont {L.~M.}\ \bibnamefont {Shekhtman}}, \bibinfo {author}
  {\bibfnamefont {M.~M.}\ \bibnamefont {Danziger}}, \bibinfo {author}
  {\bibfnamefont {J.}~\bibnamefont {Fan}}, \bibinfo {author} {\bibfnamefont
  {R.}~\bibnamefont {Du}}, \bibinfo {author} {\bibfnamefont {J.}~\bibnamefont
  {Liu}}, \bibinfo {author} {\bibfnamefont {L.}~\bibnamefont {Tian}}, \bibinfo
  {author} {\bibfnamefont {H.~E.}\ \bibnamefont {Stanley}}, \ and\ \bibinfo
  {author} {\bibfnamefont {S.}~\bibnamefont {Havlin}},\ }\href@noop {}
  {\bibfield  {journal} {\bibinfo  {journal} {Proc. Natl. Acad. Sci. U.S.A}\
  }\textbf {\bibinfo {volume} {118}},\ \bibinfo {pages} {e1922831118} (\bibinfo
  {year} {2021})}\BibitemShut {NoStop}%
\bibitem [{\citenamefont {Cohen}\ \emph {et~al.}(2002)\citenamefont {Cohen},
  \citenamefont {Havlin},\ and\ \citenamefont {ben
  Avraham}}]{cohen2002structural}%
  \BibitemOpen
  \bibfield  {author} {\bibinfo {author} {\bibfnamefont {R.}~\bibnamefont
  {Cohen}}, \bibinfo {author} {\bibfnamefont {S.}~\bibnamefont {Havlin}}, \
  and\ \bibinfo {author} {\bibfnamefont {D.}~\bibnamefont {ben Avraham}},\
  }\href@noop {} {\bibfield  {journal} {\bibinfo  {journal} {Handbook of Graphs
  and Networks: From the Genome to the Internet}\ ,\ \bibinfo {pages} {85}}
  (\bibinfo {year} {2002})}\BibitemShut {NoStop}%
\bibitem [{\citenamefont {Baxter}\ \emph {et~al.}(2011)\citenamefont {Baxter},
  \citenamefont {Dorogovtsev}, \citenamefont {Goltsev},\ and\ \citenamefont
  {Mendes}}]{baxter2011heterogeneous}%
  \BibitemOpen
  \bibfield  {author} {\bibinfo {author} {\bibfnamefont {G.~J.}\ \bibnamefont
  {Baxter}}, \bibinfo {author} {\bibfnamefont {S.~N.}\ \bibnamefont
  {Dorogovtsev}}, \bibinfo {author} {\bibfnamefont {A.~V.}\ \bibnamefont
  {Goltsev}}, \ and\ \bibinfo {author} {\bibfnamefont {J.~F.~F.}\ \bibnamefont
  {Mendes}},\ }\href@noop {} {\bibfield  {journal} {\bibinfo  {journal} {Phys.
  Rev. E}\ }\textbf {\bibinfo {volume} {83}},\ \bibinfo {pages} {051134}
  (\bibinfo {year} {2011})}\BibitemShut {NoStop}%
\bibitem [{\citenamefont {{Di Muro}}\ \emph {et~al.}(2019)\citenamefont {{Di
  Muro}}, \citenamefont {Valdez}, \citenamefont {Stanley}, \citenamefont
  {Buldyrev},\ and\ \citenamefont {Braunstein}}]{di2019insights}%
  \BibitemOpen
  \bibfield  {author} {\bibinfo {author} {\bibfnamefont {M.~A.}\ \bibnamefont
  {{Di Muro}}}, \bibinfo {author} {\bibfnamefont {L.~D.}\ \bibnamefont
  {Valdez}}, \bibinfo {author} {\bibfnamefont {H.~E.}\ \bibnamefont {Stanley}},
  \bibinfo {author} {\bibfnamefont {S.~V.}\ \bibnamefont {Buldyrev}}, \ and\
  \bibinfo {author} {\bibfnamefont {L.~A.}\ \bibnamefont {Braunstein}},\
  }\href@noop {} {\bibfield  {journal} {\bibinfo  {journal} {Phys. Rev. E}\
  }\textbf {\bibinfo {volume} {99}},\ \bibinfo {pages} {022311} (\bibinfo
  {year} {2019})}\BibitemShut {NoStop}%
\bibitem [{\citenamefont {Shang}(2014)}]{shang2014vulnerability}%
  \BibitemOpen
  \bibfield  {author} {\bibinfo {author} {\bibfnamefont {Y.}~\bibnamefont
  {Shang}},\ }\href@noop {} {\bibfield  {journal} {\bibinfo  {journal} {Phys.
  Rev. E}\ }\textbf {\bibinfo {volume} {89}},\ \bibinfo {pages} {012813}
  (\bibinfo {year} {2014})}\BibitemShut {NoStop}%
\bibitem [{\citenamefont {Valdez}\ and\ \citenamefont
  {Braunstein}(2022)}]{valdez2022emergent}%
  \BibitemOpen
  \bibfield  {author} {\bibinfo {author} {\bibfnamefont {L.~D.}\ \bibnamefont
  {Valdez}}\ and\ \bibinfo {author} {\bibfnamefont {L.}~\bibnamefont
  {Braunstein}},\ }\href@noop {} {\bibfield  {journal} {\bibinfo  {journal}
  {Physica A}\ }\textbf {\bibinfo {volume} {594}},\ \bibinfo {pages} {127057}
  (\bibinfo {year} {2022})}\BibitemShut {NoStop}%
\bibitem [{\citenamefont {Mann}\ \emph
  {et~al.}(2021{\natexlab{a}})\citenamefont {Mann}, \citenamefont {Smith},
  \citenamefont {Mitchell},\ and\ \citenamefont {Dobson}}]{mann2021random}%
  \BibitemOpen
  \bibfield  {author} {\bibinfo {author} {\bibfnamefont {P.}~\bibnamefont
  {Mann}}, \bibinfo {author} {\bibfnamefont {V.~A.}\ \bibnamefont {Smith}},
  \bibinfo {author} {\bibfnamefont {J.~B.}\ \bibnamefont {Mitchell}}, \ and\
  \bibinfo {author} {\bibfnamefont {S.}~\bibnamefont {Dobson}},\ }\href@noop {}
  {\bibfield  {journal} {\bibinfo  {journal} {Phys. Rev. E}\ }\textbf {\bibinfo
  {volume} {103}},\ \bibinfo {pages} {012309} (\bibinfo {year}
  {2021}{\natexlab{a}})}\BibitemShut {NoStop}%
\bibitem [{\citenamefont {Mann}\ \emph
  {et~al.}(2021{\natexlab{b}})\citenamefont {Mann}, \citenamefont {Smith},
  \citenamefont {Mitchell}, \citenamefont {Jefferson},\ and\ \citenamefont
  {Dobson}}]{mann2021exact}%
  \BibitemOpen
  \bibfield  {author} {\bibinfo {author} {\bibfnamefont {P.}~\bibnamefont
  {Mann}}, \bibinfo {author} {\bibfnamefont {V.~A.}\ \bibnamefont {Smith}},
  \bibinfo {author} {\bibfnamefont {J.~B.}\ \bibnamefont {Mitchell}}, \bibinfo
  {author} {\bibfnamefont {C.}~\bibnamefont {Jefferson}}, \ and\ \bibinfo
  {author} {\bibfnamefont {S.}~\bibnamefont {Dobson}},\ }\href@noop {}
  {\bibfield  {journal} {\bibinfo  {journal} {Phys. Rev. E}\ }\textbf {\bibinfo
  {volume} {104}},\ \bibinfo {pages} {024304} (\bibinfo {year}
  {2021}{\natexlab{b}})}\BibitemShut {NoStop}%
\bibitem [{\citenamefont {Allard}\ \emph {et~al.}(2012)\citenamefont {Allard},
  \citenamefont {H{\'e}bert-Dufresne}, \citenamefont {No{\"e}l}, \citenamefont
  {Marceau},\ and\ \citenamefont {Dub{\'e}}}]{allard2012bond}%
  \BibitemOpen
  \bibfield  {author} {\bibinfo {author} {\bibfnamefont {A.}~\bibnamefont
  {Allard}}, \bibinfo {author} {\bibfnamefont {L.}~\bibnamefont
  {H{\'e}bert-Dufresne}}, \bibinfo {author} {\bibfnamefont {P.-A.}\
  \bibnamefont {No{\"e}l}}, \bibinfo {author} {\bibfnamefont {V.}~\bibnamefont
  {Marceau}}, \ and\ \bibinfo {author} {\bibfnamefont {L.~J.}\ \bibnamefont
  {Dub{\'e}}},\ }\href@noop {} {\bibfield  {journal} {\bibinfo  {journal} {J.
  Phys. A}\ }\textbf {\bibinfo {volume} {45}},\ \bibinfo {pages} {405005}
  (\bibinfo {year} {2012})}\BibitemShut {NoStop}%
\bibitem [{\citenamefont {Gleeson}(2009)}]{gleeson2009bond}%
  \BibitemOpen
  \bibfield  {author} {\bibinfo {author} {\bibfnamefont {J.~P.}\ \bibnamefont
  {Gleeson}},\ }\href@noop {} {\bibfield  {journal} {\bibinfo  {journal} {Phys.
  Rev. E}\ }\textbf {\bibinfo {volume} {80}},\ \bibinfo {pages} {036107}
  (\bibinfo {year} {2009})}\BibitemShut {NoStop}%
\bibitem [{\citenamefont {Gumel}(2012)}]{gumel2012causes}%
  \BibitemOpen
  \bibfield  {author} {\bibinfo {author} {\bibfnamefont {A.~B.}\ \bibnamefont
  {Gumel}},\ }\href@noop {} {\bibfield  {journal} {\bibinfo  {journal} {J.
  Math. Anal. Appl.}\ }\textbf {\bibinfo {volume} {395}},\ \bibinfo {pages}
  {355} (\bibinfo {year} {2012})}\BibitemShut {NoStop}%
\bibitem [{\citenamefont {Ludwig}(1975)}]{ludwig1975final}%
  \BibitemOpen
  \bibfield  {author} {\bibinfo {author} {\bibfnamefont {D.}~\bibnamefont
  {Ludwig}},\ }\href@noop {} {\bibfield  {journal} {\bibinfo  {journal} {Math.
  Biosci.}\ }\textbf {\bibinfo {volume} {23}},\ \bibinfo {pages} {33} (\bibinfo
  {year} {1975})}\BibitemShut {NoStop}%
\bibitem [{Note1()}]{Note1}%
  \BibitemOpen
  \bibinfo {note} {Here, we use the principal branch of the Lambert
  function.}\BibitemShut {Stop}%
\bibitem [{\citenamefont {Ferreira}\ \emph {et~al.}(2012)\citenamefont
  {Ferreira}, \citenamefont {Castellano},\ and\ \citenamefont
  {Pastor-Satorras}}]{ferreira2012epidemic}%
  \BibitemOpen
  \bibfield  {author} {\bibinfo {author} {\bibfnamefont {S.~C.}\ \bibnamefont
  {Ferreira}}, \bibinfo {author} {\bibfnamefont {C.}~\bibnamefont
  {Castellano}}, \ and\ \bibinfo {author} {\bibfnamefont {R.}~\bibnamefont
  {Pastor-Satorras}},\ }\href@noop {} {\bibfield  {journal} {\bibinfo
  {journal} {Phys. Rev. E}\ }\textbf {\bibinfo {volume} {86}},\ \bibinfo
  {pages} {041125} (\bibinfo {year} {2012})}\BibitemShut {NoStop}%
\end{thebibliography}%

\end{document}